\documentclass[a4paper,10pt]{article}

\usepackage[english]{babel}
\usepackage[latin1]{inputenc}
\usepackage{graphicx}
\usepackage{epsfig}
\usepackage{hyperref}
\usepackage{wrapfig}

\usepackage[normalem]{ulem}

\newcommand{\pr}{\partial}
\newcommand{\rta}{\rightarrow}

\newcommand{\p}{\prime}

\newcommand{\ra}{\rangle}
\newcommand{\la}{\langle}

\newcommand{\mf}{\mathbf}

\newcommand{\bq}{\mf{q}}

\newcommand{\bn}{\mf{n}}
\newcommand{\fis}{\Bigg\lceil_\sigma}

\newcommand{\fiszero}{\Bigg\lceil_{\sigma_0}}

\newcommand{\fisone}{\Bigg\lceil_{\sigma_1}}

\title{Thermodynamical Phase transitions, the mean-field theories,
and the renormalization (semi)group: A pedagogical introduction}
\author{Navinder Singh\footnote{navinder.phy@gmail.com}}

\begin{document}

\maketitle
``Since 'tis Nature's law to change,\\
\indent Constancy alone is strange'' ---  Earl of
Rochester.
\begin{abstract}
While analyzing second order thermodynamical phase transitions, Lev Landau (the famous Russian
physicist) introduced a very vital concept, the concept of an "order parameter". This not only
amalgamated the previous fragmentary theoretical understanding of phase transitions (an arsenal of
mean-field theories) but also it put forward the important theory of
"spontaneous symmetry breaking". Today, order parameter concept is a
paradigm both in condensed matter physics and in high energy physics,  and
Landau theory is a pinnacle of all mean-field theories.  Mean field
theories are good qualitative descriptors of the phase transition
behavior. But all mean-field theories (including Landau's theory) fail at
the critical point (the problem of large correlation length). The problems
with large correlation length in quantum many-body systems are the hardest
problems known in theoretical physics (both in condensed matter and in
particle physics). It was Ken Wilson's physical insights and his powerful
mathematical skills that opened a way to the solution of such hard
problems.

This manuscript is a perspective on these issues. Starting with
simple examples of phase transitions (like ice/water; diamond/graphite
etc.)  we  address the following important questions: Why does
non-analyticity (sharp phase transitions) arise when thermodynamical
functions (i.e., free energies etc) are good analytic functions? How does
Landau's program unify all the previous mean-field theories? Why do all
the mean-field theories fail near the critical point? How does Wilson's
program go beyond all the mean-field theories? What is the origin emergence and universality?
\end{abstract}

\tableofcontents


\section{Introduction to Thermodynamical Phase Transitions}
Mankind is familiar with various phases of matter  and transitions between the phases from
antiquity. For example, ice-water-steam; ice
is solid that exhibit characteristic rigidity, water is liquid and it flows like other liquids, and
steam is a gas, much like other gases. It is only in the 19th century that with the advent of the
science of thermodynamics the phases of matter and their transformations were rationalized within
the scientific domain. The phase transition of water to steam gave birth to the great discovery of the
steam engine by Thomas Newcomen in 1712, and it resulted in the subsequent great industrial
revolution first in Britain and then in Europe. It is beyond doubt that man is able to use, for the
service of humanity, the great practical potential of the fundamental understanding of the
materials and their transformations.

Aim of the the present paper is to give a broad overview of the phase transition phenomena from
fundamental perspective starting
from early works of Andrews and van der Waals. The Mean-Field Theories (MFTs) were developed to understand thermodynamical phase transitions from atomic/molecular perspective, and van der Waals was the pioneer in this\cite{23}.  We discuss various Mean-Field Theories (MFTs)
and their reformulation by Lev Landau in 1937\cite{1}. MFTs are good qualitative descriptors of the phase transition
behavior. But all mean-field theories fail at
the critical point including the Landau's formulation. At the critical point correlation length
(defined precisely in the text) diverges and fluctuations dominate over the average behavior. It is
this failure of the MFTs that triggered further investigations that resulted the formulation of renormalization group by Wilson and others in 1970s\cite{2}. The problems
with large correlation length in quantum many-body systems are the hardest
problems known in theoretical physics (both in condensed matter and in
particle physics) and this branch of physics is far from being closed. 

The present paper is a pedagogical introduction to this vast but coherent topic. The story of the successes and failures of MFTs and their systematic replacement at the critical point by the renormalization group theory is given in a clear and pedagogical way. We also address, more
specifically, the following questions: Why does
non-analyticity (sharp phase transitions) arise when thermodynamical
functions (i.e., free energies etc) are good analytic functions? How does
Landau's program unify all the previous mean-field theories? Why do all
the mean-field theories fail near the critical point? How does Wilson's
program go beyond all the mean-field theories? What is origin emergence and universality?

The answers of above questions are well known\cite{2,3,4}. Presentation here is sufficiently self-contained to address the above questions. {\it Thus it will be useful to the students learning the subject.}  We also express our standpoint regarding the failure of MFTs, and point out that MFTs faithfully reproduce diverging fluctuations at the critical point and divergence of fluctuations can be understood most clearly by recognizing the fact that the property of Statistical Independence (SI) is violated in systems with long correlation length (when various parts of the system become statistically correlated)\cite{5}. And the fundamental thesis of
Boltzmann-Gibbs statistical mechanics--fluctuations converge as $\propto \frac{1}{N}$ in
sum-function observables\cite{5}--is no longer applicable in systems with large correlation length
(of the order of the system's size) making the fluctuations to dominate over the average behavior. The reason why MFTs predict wrong values of critical indices is given in\cite{6}, that also clarifies a common misconception.

The paper is organized as follows. In the next subsection we give some examples of phase transitions. The next section (section 2) is devoted to Ehrenfest's classification and to some mean-field theories. Then in section 3 we present Landau's formulation of MFTs. Section 4 deals with the problems of MFTs. After recognizing the difficulties when dealing with problems  with long correlation length we present Kadanoff's construction and Wilson's Renormalization Group (RG) formulation that rectify the problems of MFTs (section 5). This is illustrated by a specific example of $\phi^4$-theory. Although there are many different formulations of renormalization group method and has been applied to many different problems\cite{2,3,4,7,8,9,10}, our point here is to present the essence of the method for pedagogical purposes.  In section 6 we also address the important question: Why do all
the mean-field theories fail near the critical point and how does Wilson's program go beyond MFTs? A short biographical sketch of Ken Wilson  is given in the appendix.

\subsection{Some examples}
Phase transitions are ubiquitous in Nature! Most common phase transition is that of water and ice
(Figure 1(a)). We see this when we put liquid water in the refrigerator for making ice, a common experience.
We know that to make ice from water we need to extract some heat for a given mass of water. This is
called the latent heat and for water it is $334 ~Joules/gram$ . Thus to make ice from water you have
to extract this much amount of heat per gram from water (i.e., by cooling it)  and to convert back water
from ice  you have to supply that amount (i.e., by heating the ice). You also note that when you
cool
water to convert it into ice, it expands, i.e., the volume of ice for a given mass of water is more
than the volume of liquid water of that mass, or ice floats on water. These facts can be quantitatively 
described by introducing the thermodynamical free energy\cite{20}. Discontinuity of its first order
derivatives with respect to thermodynamical parameters leads to above discontinuities. We will
explain this in the section devoted to Ehrenfest's classification. 
\begin{figure}[h!]
\centering
\begin{tabular}{cc}
\includegraphics[height = 5cm, width =5.5cm]{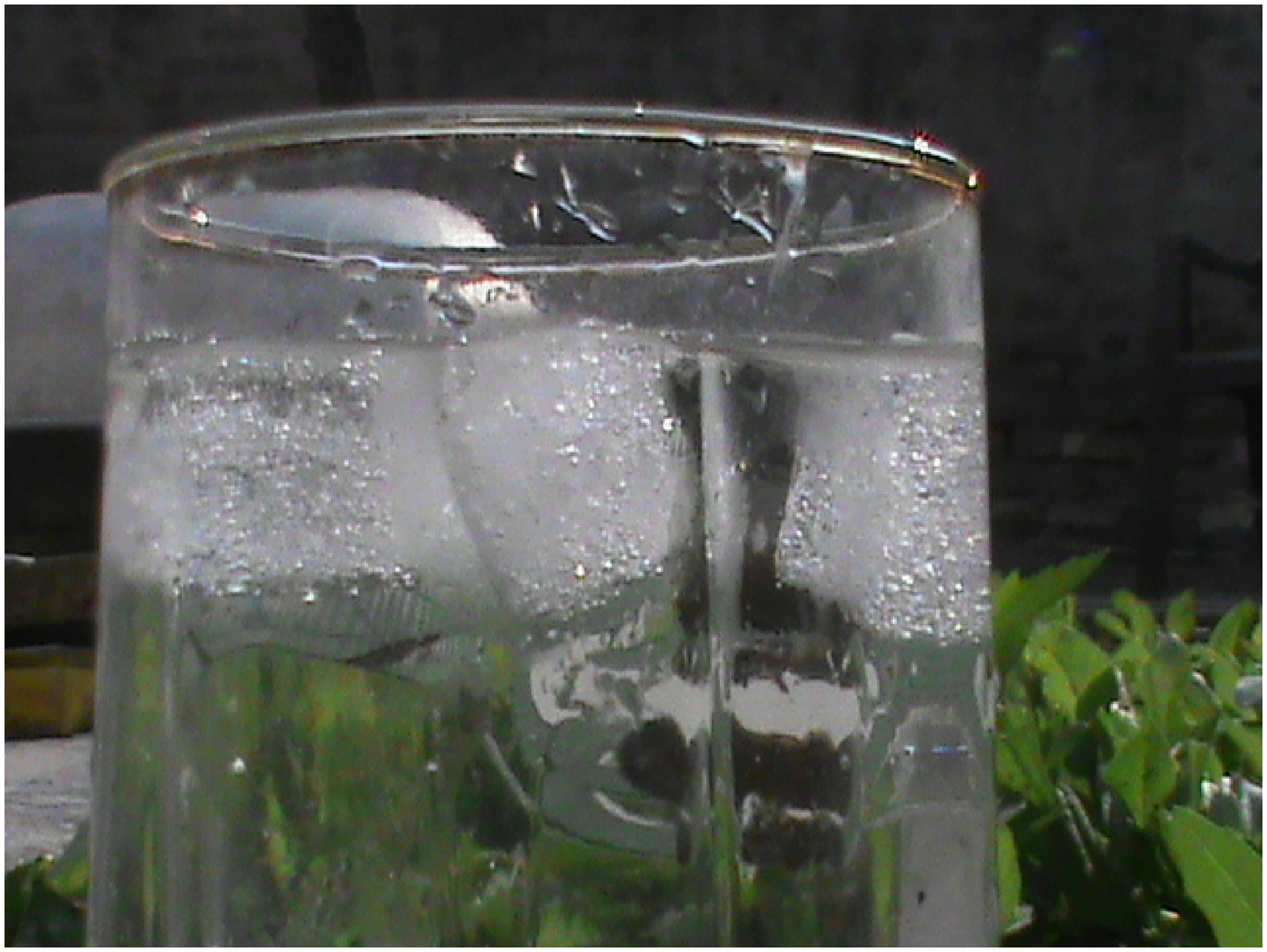}&
\includegraphics[height = 5cm, width =6cm]{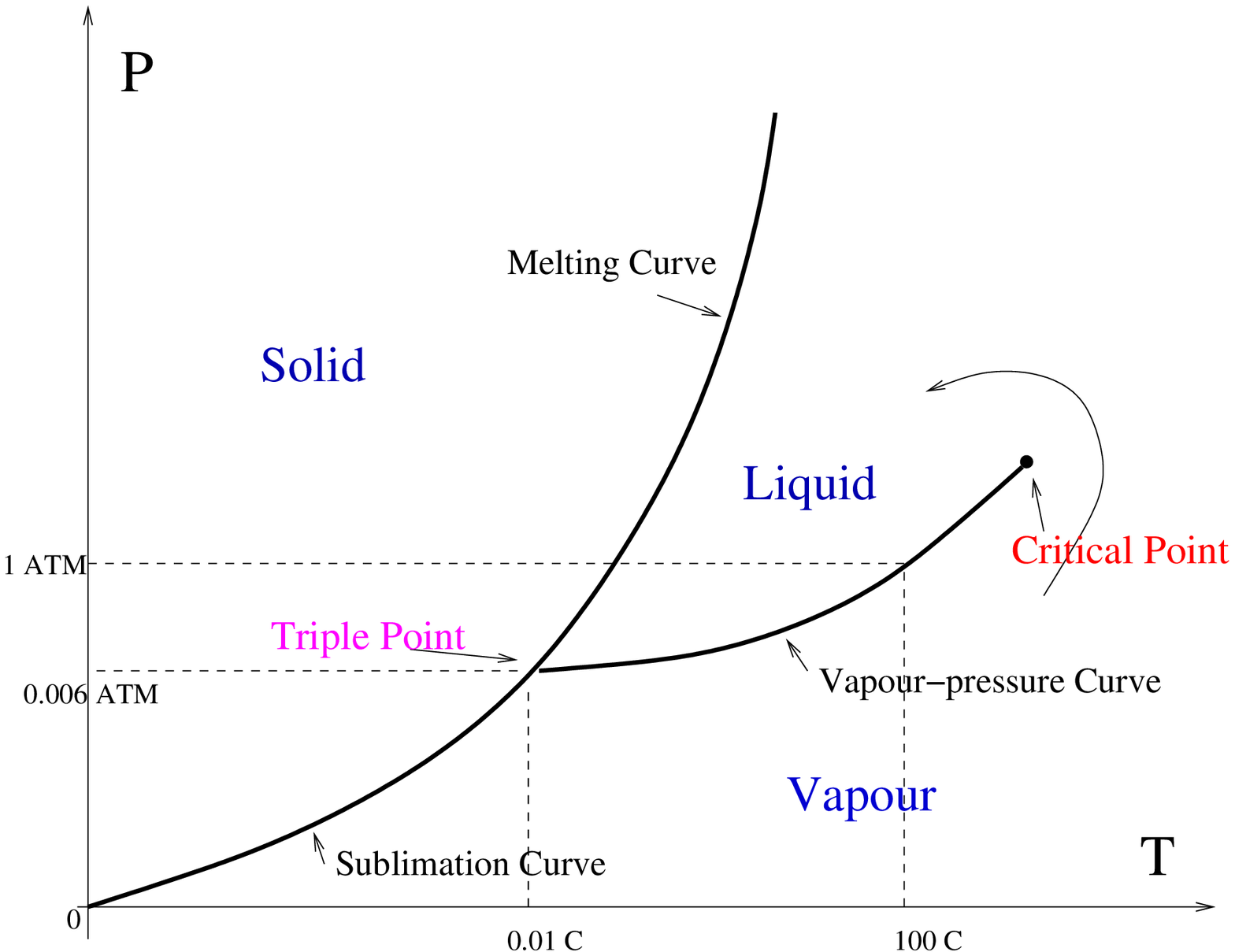}\\
(a)&(b)
\end{tabular}
\caption{ (a) Three phases of water. (b) Phase diagram of a simple substance like water.}
\label{water}
\end{figure}


Figure~\ref{water}(b) depicts a typical phase diagram of a ``simple'' substance like water for
example
(only change one has to make is that of a negative slope of the liquid-solid phase line for the
case of water).  Phase diagram depicts
the kind of phases that are taken by a given substance under a given set of external conditions.
These
external conditions are the thermodynamical parameters of the system, for example, Pressure (P) and
Temperature (T) for a simple substance, or, temperature and external magnetic field for a magnetic
substance, like, for example, a ferromagnetic.

The phase diagram, as in figure 1(b), is  a two dimensional projection on the P-T
plane of a three dimensional picture, with free energy along an axis coming out of the plane of
the paper (for example, referring to figure ~\ref{water}(b)) and perpendicular to the P-T axis. So in
this 3-D
picture one has three free energy {\it surfaces.} A substance is stable in a given region of the
phase
diagram if its {\it free energy surface} takes the minimum value. Consider the gaseous regime, in this, the
free energy surface for the gaseous phase takes the minimum value as compared to two other free
energy surfaces that of liquid and solid. The phase boundary lines shown in figure \ref{water}(b)
are the
lines at which above mentioned surfaces cut each other. And at each phase line free energies of two
substances separated by that line takes the same value. Thus on each phase line both can co-exist.
The point at which all can co-exist is called the triple point.  Another important point is called
the critical point. It is a point in a phase diagram where first order phase transition goes to
second order phase transition. The discontinuity in the first order derivatives of $g$ (latent heat
and change of volume per unit mass) goes to zero. Thus in liquid-gas transition the density
difference (between gas and liquid) becomes zero and it is not possible to distinguish gas from
liquid at the critical point\footnote{But density fluctuations dominate over all length
scales at the critical point. Density fluctuations on the length scale of the wavelength of visible light cause strong
scattering of light and results in a very spectacular phenomenon called the critical  opalescence.
Here is a quick experiment\cite{11}. Take equal quantities of cyclohexane and acetic anhydride in a test
tube. Two  liquids do not mix at room temperature, but on heating they mix. If one let the
mixture to cool, near $~53$ C~ the mixture turns milky white----A spectacular phenomenon called critical
opalescence.}. However, the second order derivatives of the Gibbs free energy remains discontinuous
(see the section on Ehrenfest's classification).

\begin{figure}[h!]
\centering
\begin{tabular}{cc}
\includegraphics[height = 6cm, width =6cm]{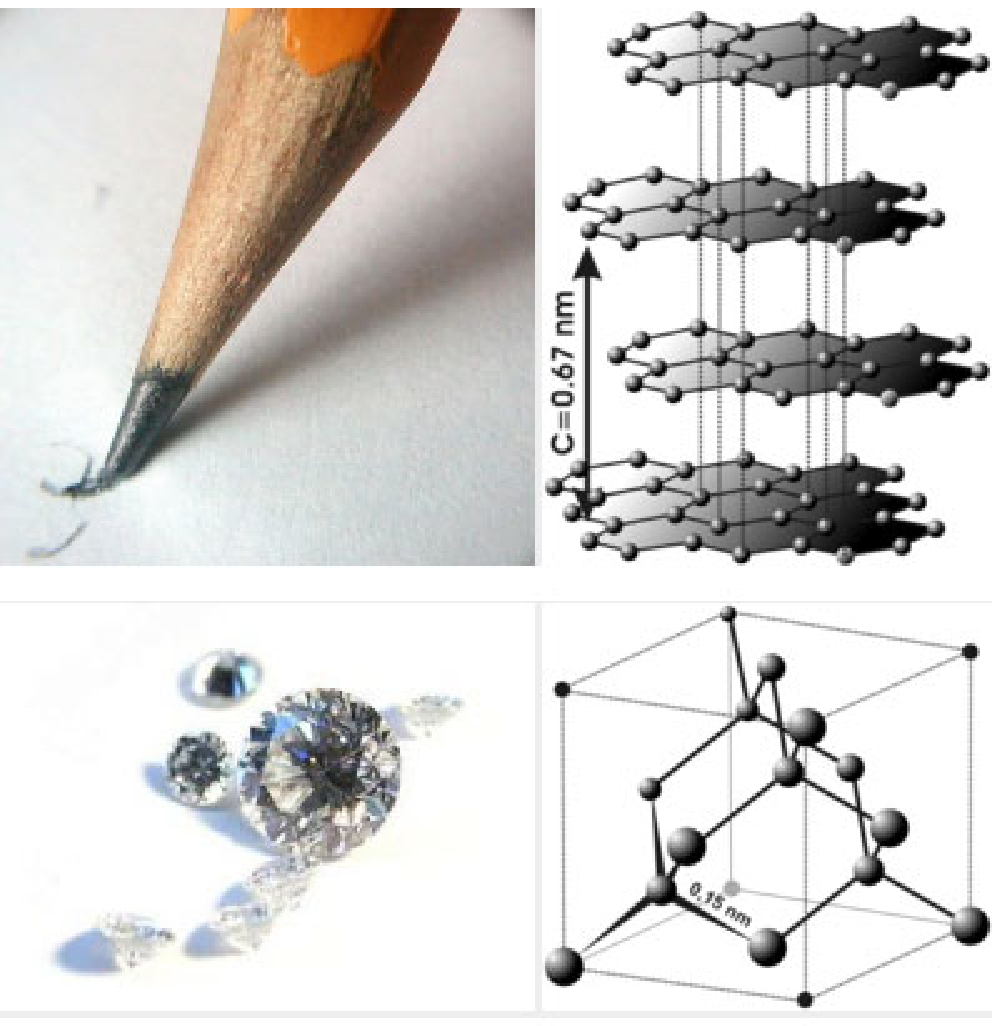}&
\includegraphics[height = 6cm, width =6cm]{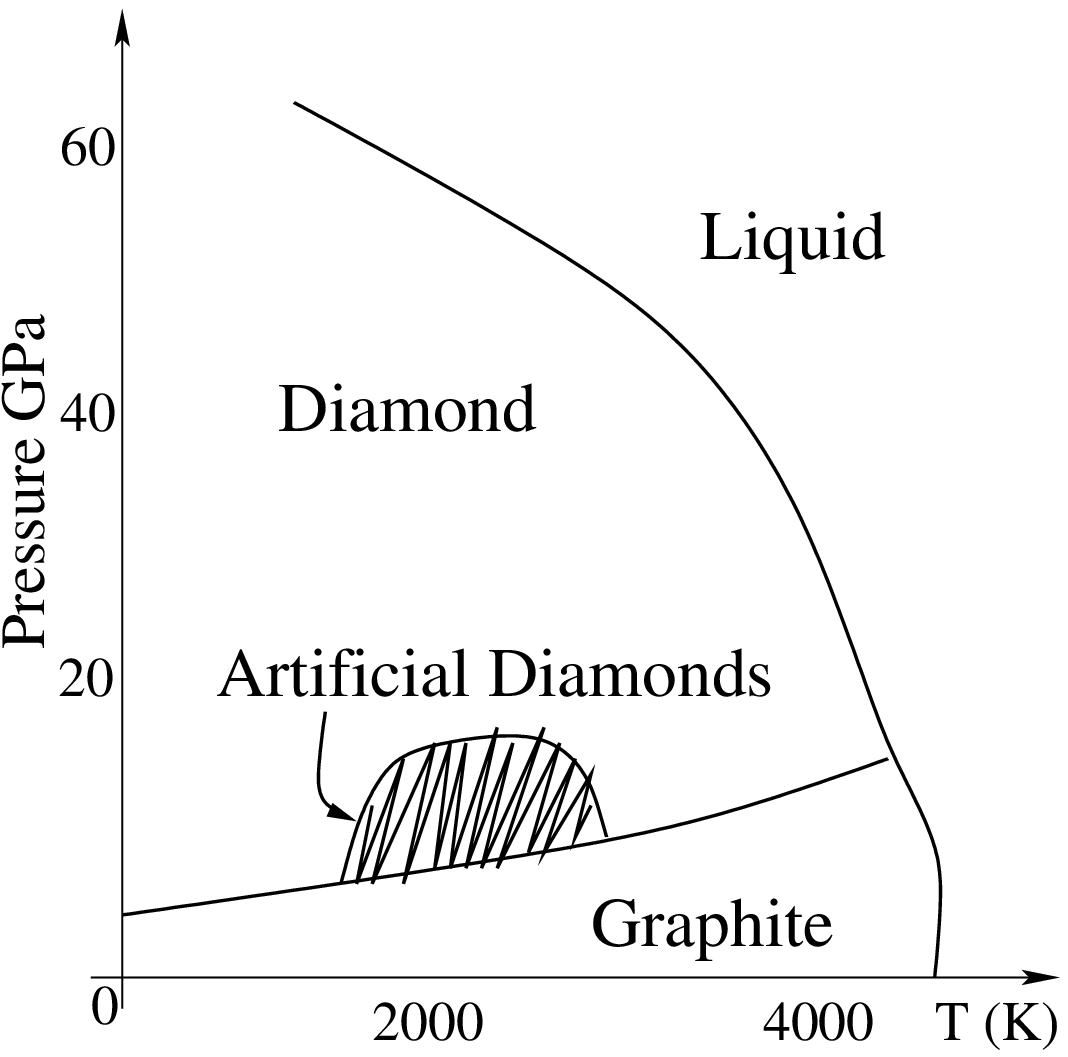}\\
(a)&(b)
\end{tabular}
\caption{(a) Allotropes of Carbon: Graphite is soft! while Diamond is the hardest material known (Figure source: http://mrsec.wisc.edu/Edetc/nanoquest/carbon/index.html).
(b) Phase diagram of Graphite.}
\label{diam}
\end{figure}

Diamond and Graphite: Another example of phase transitions is that of diamond and Graphite (figure~ (\ref{diam})). These
are two allotropes of  Carbon. Pencils write because layers of graphite easily slip past each other
and can coat another object, such as paper. Diamonds look very very different from pencils! They are
the hardest mineral
known!! This is an example of the structural phase transition in which Carbon atoms rearrange
themselves. Diamonds and Graphite are naturally occurring substances. Diamonds formed from Carbon
deep under earth ($\sim 200~km$) by geologic processes under larger pressure and temperature
environments (as compared to ambient conditions). One can even transform Graphite into Diamonds--in
a Lab! This is possible by applying a very large pressure, $100,000$ times that of normal
atmospheric pressure, with graphite sample kept at $2000~$K (figure \ref{diam} (b)). Thus formed artificial diamonds are
smaller in size as compared to diamonds created by geological processes.  Nature is a smarter
creator! One would like to say!

Superfluidity and Superconductivity: Phase transitions can also lead to quantum behavior at
macro-scale.
Superconductivity and superfluidity are two prime examples\cite{12}. One interesting example is that of
Helium. Helium exists in nature in two stable isotopes. One common isotope is $_2 He^4$. The other
rare isotope $_2 He^3$ is one atom for a million atoms of $_2 He^4$ in earth's atmosphere. 
\begin{figure}
\centering
\begin{tabular}{cc}
\includegraphics[height = 5cm, width =6cm]{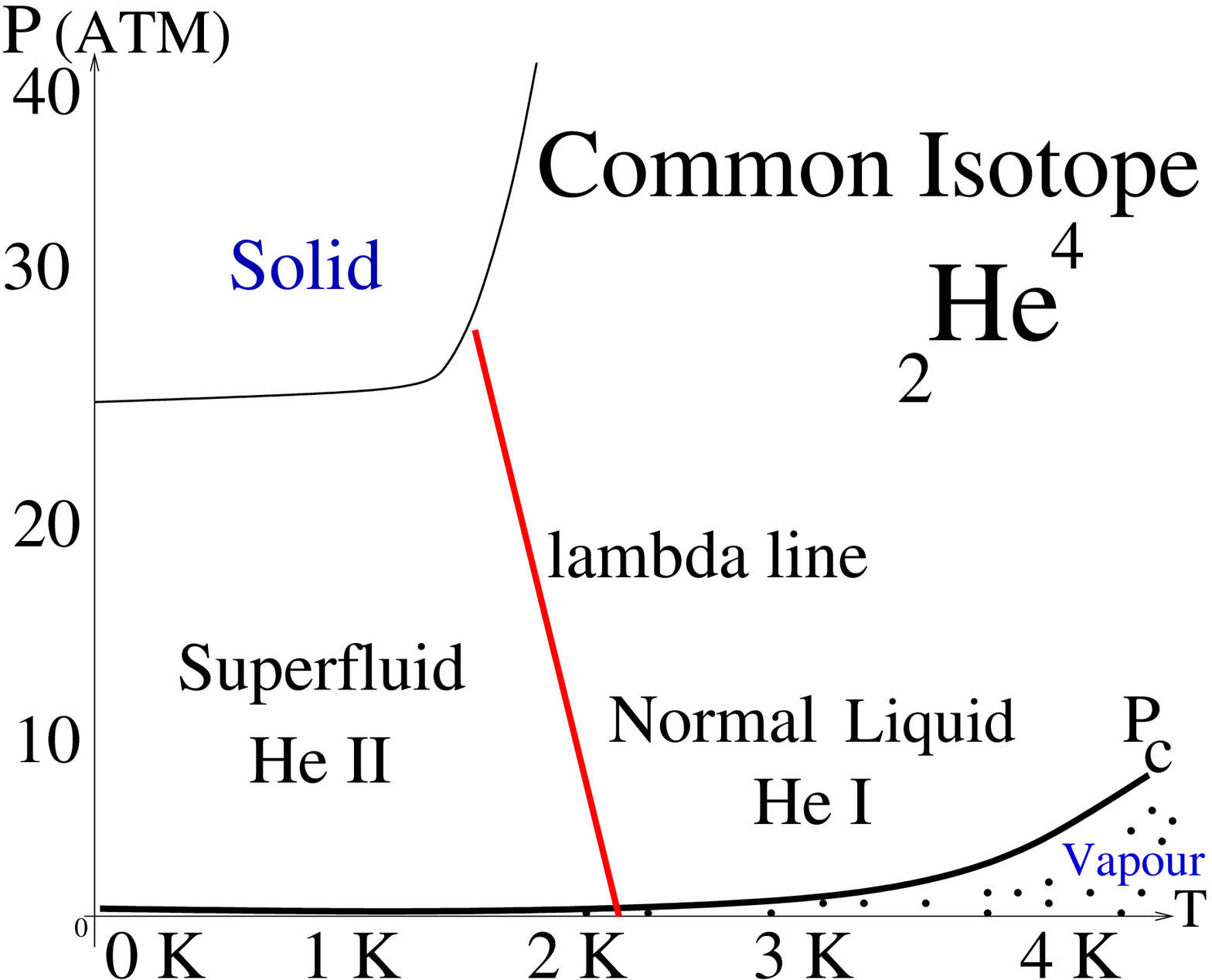}&
\includegraphics[height = 5cm, width =5.5cm]{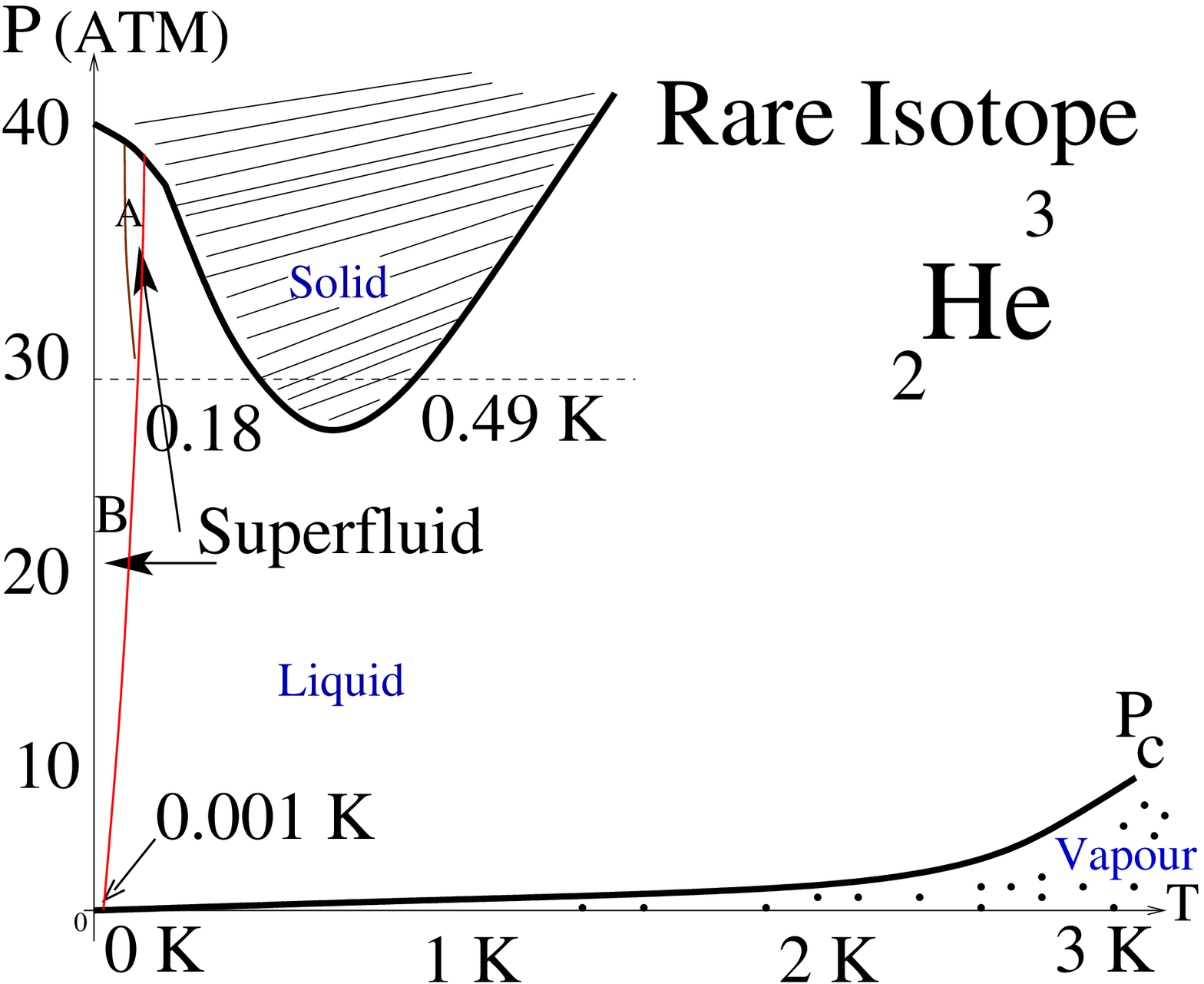}\\
(a)&(b)
\end{tabular}
\caption{Phase transitions in Helium. (a) Common Isotope $_2 He^4$, (b) The rare isotope $_2 He^3$.}
\label{he}
\end{figure}
Consider first the phase diagram of common isotope $_2 He^4$ (Figure ~\ref{he}(a)). It liquefies at
$4.2~K$ at the ambient
pressure (Heike Kamerlingh Onnes was the first  to liquefy helium. He liquefied it in 1908
and later discovered superconductivity in Mercury in 1911). When cooled further $\sim 2.17~K$ it
transforms to a new form of liquid (a new thermodynamical phase called the superfluid phase). In
this phase liquid Helium flows without friction (zero viscosity) and becomes a
``super-heat-conductor''. This superfluid phase of $_2 He^4$ is also called the HeII phase (the
normal one is called HeI). Heat capacity shows an infinite discontinuity at $2.17 ~K$ (figure
(\ref{lam})). There
is no latent heat involved\footnote{But it is not the second order phase transition! see
Ehrenfest's classification in the next section.}. The shape of
heat capacity curve resembles with that of a Greek letter lambda ($\lambda$). Thus it is also called
the lambda-transition in liquid Helium literature. Liquid Helium can be made solid by applying a
pressure of 25 atm for temperature less than 2 K. Higher temperatures require higher pressure for
solidification (figure \ref{he} (a)). Interestingly it does not have a triple point!
\begin{figure}[h!]
\includegraphics[height = 6cm, width =8cm]{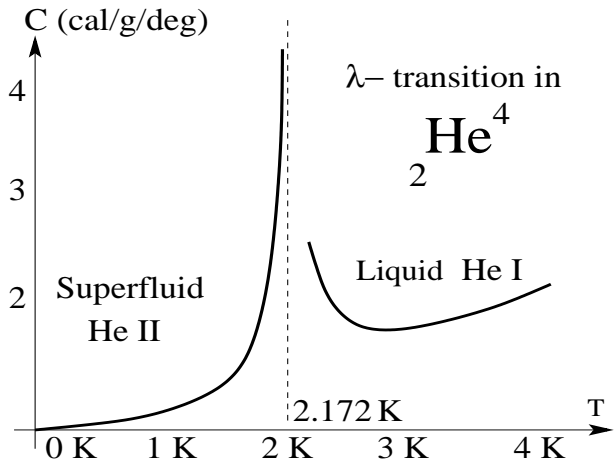}
\caption{The lambda-transition in common Helium.}
\label{lam}
\end{figure}
The rare isitope $_2 He^3$ behaves exotically (figure \ref{he}(b)). If you start from a very low
temperature, say $0.0001~ K$, with a sample of $_2 He^3$ at 30 atm pressure, you first have liquid
Helium. On reaching $0.18~K$ it solidifies! It remains solid upto $0.49~K$ and again turns liquid
for higher temperatures. This behavior of the rare isotope of Helium is quite unusual (as common
solids remain solid with reducing temperature!).

The superfluidity of the normal isotope $_2 He^4$ is due to the Bose-Einstein condensation below
the condensation temperature $\sim 2.17~K$. As $_2 He^4$ atoms are bosons they condense into the
ground state with order parameter (the condensate wavefunction) as $\rho e^\phi$. Where $\rho$ is
the superfluid density and $\phi$ is phase of the wave function. All the observed properties can be
understood using two-fluid picture (see for details\cite{13}). The reason why Helium remains liquid down to
the lowest temperatures achievable at normal pressure is the large amplitudes of the zero-point
quantum oscillations that disrupt the crystalline order.

The case of the rare isotope is much more interesting. As $_2 He^3$ atoms are fermions they cannot
show BEC directly and thus superfuildity. But experimentally superfuildity has been seen. The
reason for superfluidity in the rare isotope is the formation of Cooper pairs between Helium atoms
much like the BCS (Bardeen-Cooper-Schrieffer) theory of superconductivity\cite{14}, but with important differences: As there is no
crystalline lattice, phonons are not operative in forming Cooper pairs (in contrast to simple
metals explainable by BCS theory). The origin of Cooper pairing in the rare isotope is magnetic and
Coopers pairs resides in higher angular momentum states with internal degrees-of-freedom (see
Anderson-Morel\cite{15}).

Superconductivity is an another example that show macroscopic quantum effects. First discovered by
Heike Kamerlingh Onnes in 1911, it took about fifty years of
theoretical efforts and finally the microscopic understanding came with BCS theory of superconductivity in 1957\cite{14}. 
\begin{figure}
\centering
\begin{tabular}{cc}
\includegraphics[height = 5cm, width =6cm]{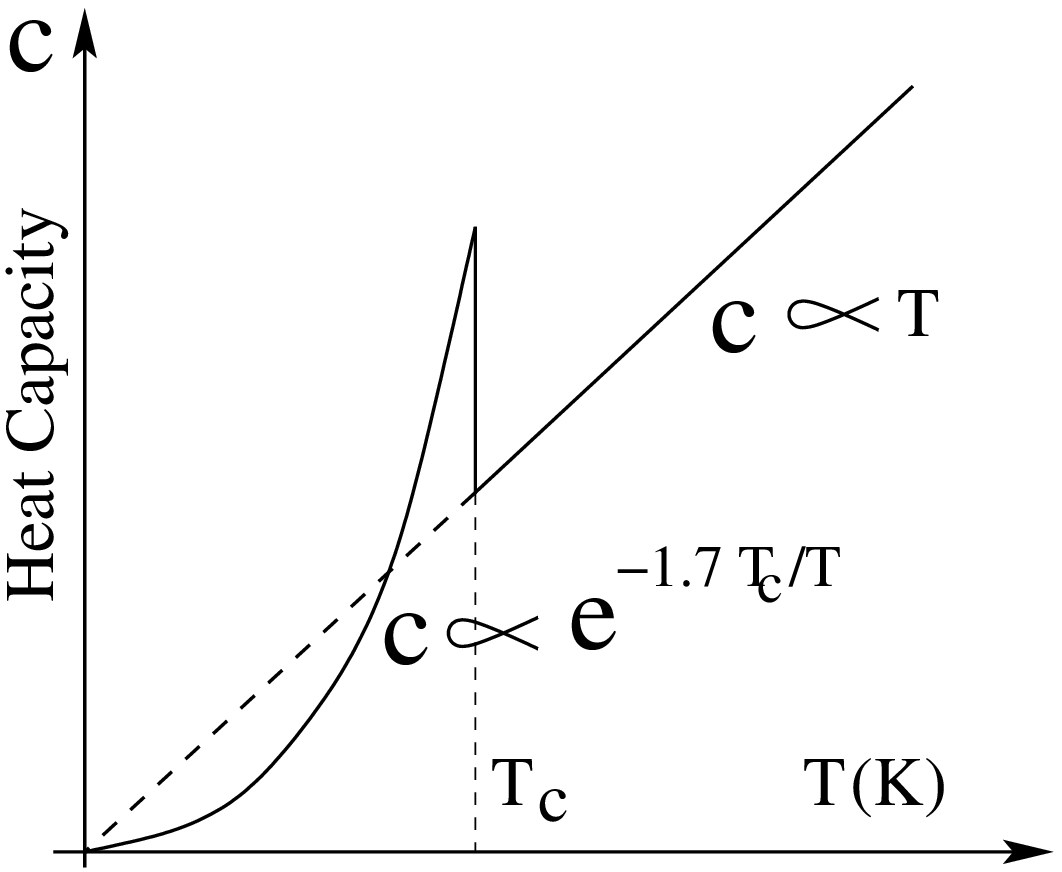}&
\includegraphics[height = 5cm, width =6cm]{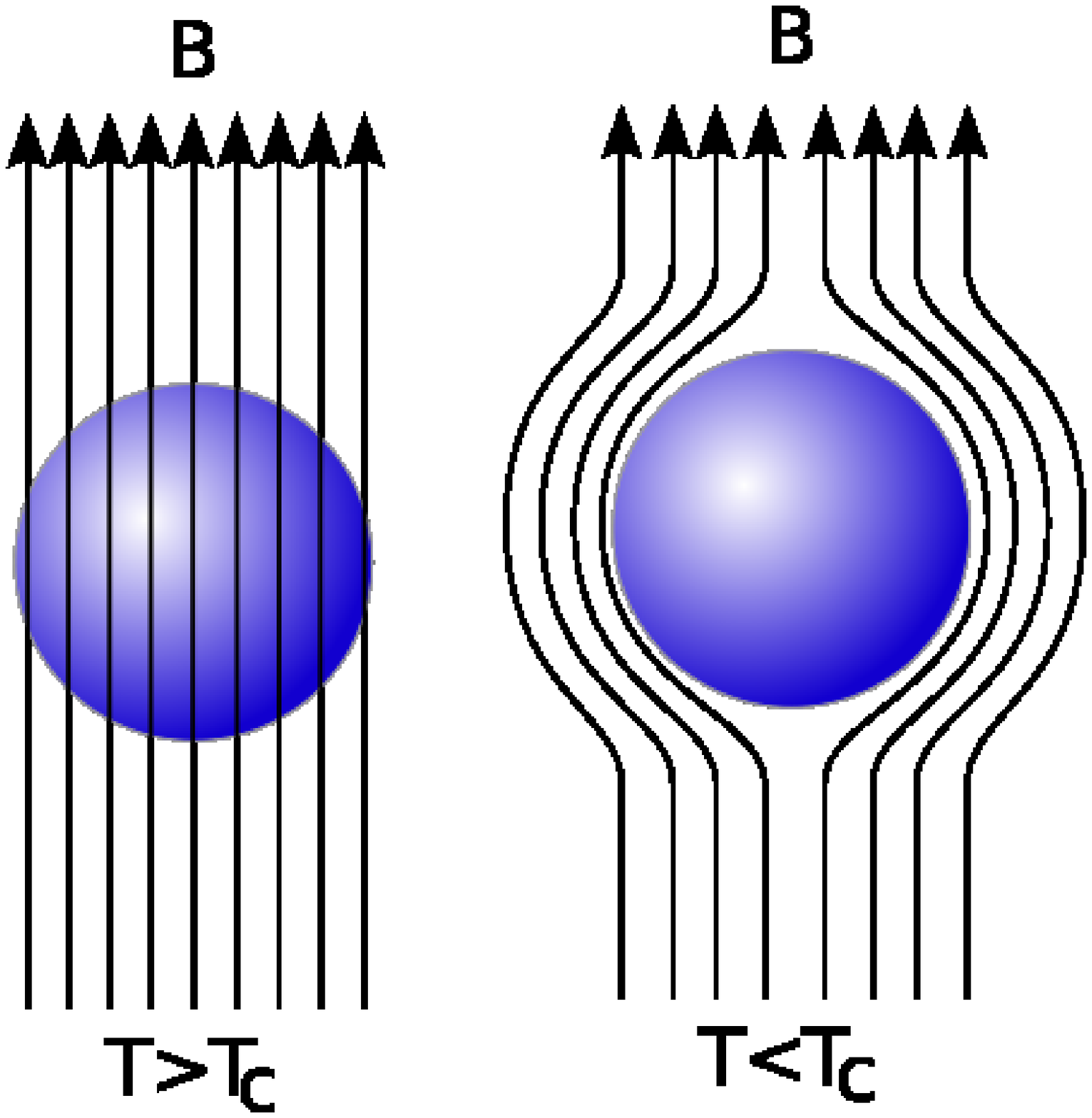}
\end{tabular}
\caption{Superconducting phase transition. (a) Heat capacity show a finite discontinuity at $T_c$,
(b) Superconducting substances show Meissner effect (Image: public domain/Wikimedia Commons.).}
\end{figure}
Loosely speaking the basis of superconductivity is the Bose-Einstein condensation of Cooper pairs
formed due to mutual attractive interaction mediated by the exchange of lattice phonons. The
important difference between rare isotope of Helium and superconductivity in simple metals is that
the size of a Cooper pair in metals is much greater than mean particle separation ($\sim 1000$ times the lattice constant)
while in Helium it is like forming a bosonic molecule by two fermionic atoms (the size of the
Cooper pair is $~$ mean particle separation). BCS theory quantitatively account for many observed
properties of simple superconductors (excluding Cuprate superconductors\cite{12,16}).

Magnetism of Iron: Iron loses its magnetic properties above $~770~C$ (the Curie
temperature).  What happens is that due to thermal agitation the ordered arrangement of electronic
spins becomes unstable. The origin of magnetism in Iron is due to a dual role played by d-band
electrons\cite{17,17a}. They participate in metallic bounding as well as act like localized spins. Iron drives
its magnetism from electron-electron repulsive Coloumb interaction due to which they avoid each
other in real space and that is effectively executed by keeping spins in the same direction. This
a rough picture (see for more accurate and quantitative analysis\cite{17a}).

Cosmological Phase Transition: The origin of SU3, Isospin, and Parity! It has been postulated (after the discovery of Hubble expansion) that when our universe was very
young (only sub-pico seconds of age) after the Big-Bang, our universe underwent a phase
transition from that high energy scale--called the Cosmological phase transition. Below that high
energy
scale the electroweak symmetry is broken. This the phase transition of vacuum itself in which
vacuum expectation value of a scalar field takes non-zero value\cite{18}. This is intimately connected to the
Higgs mechanism\cite{19}.

The above are some examples of phase transitions. Obviously, there are many more examples. The above
are sufficient for our purpose of illustrating the mean-field theories and renormalization group.
Before we discuss the MFTs we first review the work of Ehrenfest on the classifications of phase
transitions.

\section{Phase transitions without Landau's paradigm}

\subsection{Ehrenfest's classification of phase transitions}
Paul Ehrenfest (1880-1933)\footnote{He was a student of Ludwig Boltzmann.} classified various phase
transitions in 1933 according to the behavior of the derivatives of the Gibbs free energy
(unfortunately, later that year he killed himself as he was suffering from severe depression).
\begin{wrapfigure}{r}{0.3\textwidth}
  \vspace{-20pt}
  \begin{center}
    \includegraphics[width=0.3\textwidth,height=3cm]{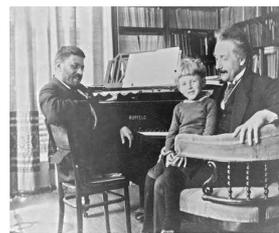}
  \end{center}
  \vspace{-15pt}
\caption{Ehrenfest and his son with Einstein $~1920$.}
  \vspace{-10pt}
\end{wrapfigure}
According to this scheme, the order of a phase transition is given by the order of the
discontinuous derivative
of Gibbs free energy. On the line separating the two phases (say, in the
liquid-gas transition) the Gibbs free energies per unit mass $g(T,P)$ of the two phases take the
same value. On the left of the line, $g$ is less for the liquid, and on the right of it it is less
for the gas. By definition\cite{20}:
$g(T,P)= U-Ts+Pv$ or $dg = -s dT + v dP$ with
\[Entropy/mass~=~s= -\frac{\pr g}{\pr T}|_{P}.\]
\[Volume/mass~=~v= \frac{\pr g}{\pr P}|_{T}\]
First order phase transitions are those in which first order derivatives of the Gibbs' Free energy
are discontinuous: i.e., if the above derivatives are discontinuous.
If we cross the liquid-solid line vertically in up direction (at constant temperature, see figure
~\ref{water} (b)) clearly:
\[\frac{\pr g_l}{\pr P}|_T > \frac{\pr g_s}{\pr P}|_T\]
As the system goes from liquid stability to solid stability (an experimental fact!).
This implies that:
\[v_l =\frac{\pr g_l}{\pr P}|_T  > v_s =\frac{\pr g_s}{\pr P}|_T \]
In other words, for a given mass, volume of the liquid phase is more than that of the
solid phase, or, substances  expand on melting!\footnote{Note that the case of water is opposite (ice is lighter) as the ice-water (solid-liquid) line has negative slop.}

If we cross the liquid-gas line horizontally (ref. figure~\ref{water}(b)) from left to right (at
constant pressure) clearly:
\[\frac{\pr g_s}{\pr T}|_P > \frac{\pr g_l}{\pr T}|_P\]
As system goes from solid stability to liquid stability (again an experimental fact!). Above implies
that 
\[-s_s>-s_l\]
or
\[s_s<s_l,~~~\Rightarrow \frac{q_l-q_s}{T}>0.\]
Where $q_l-q_s=l$ is called the latent heat of solid-to-liquid transition, which is positive! as it
should be! Thus heat must be absorbed from solid to liquid transition. Examples of first order are:
Solid-liquid-vapor, transitions in simple fluids,  allotropic transformations (like diamond graphite)
etc.

For the second order phase transitions, first order derivative of Gibbs free
energy, i.e., $v,~s$, are continuous,
but the second order derivatives, i.e., heat capacity $c$, and
compressibility $\kappa$ are discontinuous. Thus no latent heat and volume change occur during a
second order phase transition. But in a second order transition, heat capacity and compressibility
show {\it finite} discontinuity. An ideal second order
transition is the superconducting transition in zero magnetic field\cite{20}.

Phase transition in liquid Helium is not second order (as the discontinuity in heat capacity
is infinite). This is a special transition called the $\lambda-$transition.

Above is called the Ehrenfest's classification scheme. One can define higher order
phase transitions along these lines (third  and higher order derivatives of the Gibbs free energy), but those
are seldom observed in practice. Ehrenfest's scheme is based on the fundamental thermodynamical
principle: the phase of minimum free energy is the stable one\cite{20}. From this it derives other
properties of phase transitions like latent heat, volume changes etc. Importantly, it does not
provide more fundamental statistical mechanical formulation of the phase transition behavior.  We
will see in the next section that MFTs are the first steps in these directions. 

We will also see that with the Landau's paradigm above scheme of Ehrenfests' can be nicely
re-interpreted by using the order parameter concept.

\subsection{The Mean-Field Theories(MFTs)}

The above is the macroscopic way of understanding/classifying the phase transitions {without any
microscipic (atomic/molecular) point of view}. Attempts to understand the laws of heat from the
microscipic (atomic/molecular) perspective resulted in the science of statistical
mechanics and Ludwig Boltzamnn (1844-1906) was the pioneer\cite{21}. Similar to this philosophy, to
understand phase transition phenomena
from atomic/moleculaer perspective, early attempts were made by van der
Waals\footnote{The thought process of understanding macro-phenomema from microscopic perspective is
very old one, and is related to one of the very basic aspect of human psychology, namely the habit of understanding phenomena around from the basic ``building blocks''.  Examples go back to BC era, for example, Democritus was the
first to postulate atoms and atomic constitution of matter. Other example is that of statistical mechanics (thermodynamics from atomic/molecular perspective) and so on. And this unique habit of mankind resulted in a body of organized
knowledge that we call science. Author do not want to go into the debates of reductionism and emergence philosophies\cite{22}}.

Thus the credit of first mean-field theory goes to van der Waals'\cite{23}. Van der Waals was interested in
the notion of continuity of matter, namely, how a gaseous substance, on cooling, transforms into a
liquid while chemically, at molecular level things remain intact\footnote{There is no chemical
reaction when water converts from liquid to steam and vice versa.}. His main concern was the ideal gas equation
and the problems of the Boyle's law.  Here is a rough sketch of his argument.

\begin{figure}
\centering
\begin{tabular}{cc}
\includegraphics[height = 4cm, width =5cm]{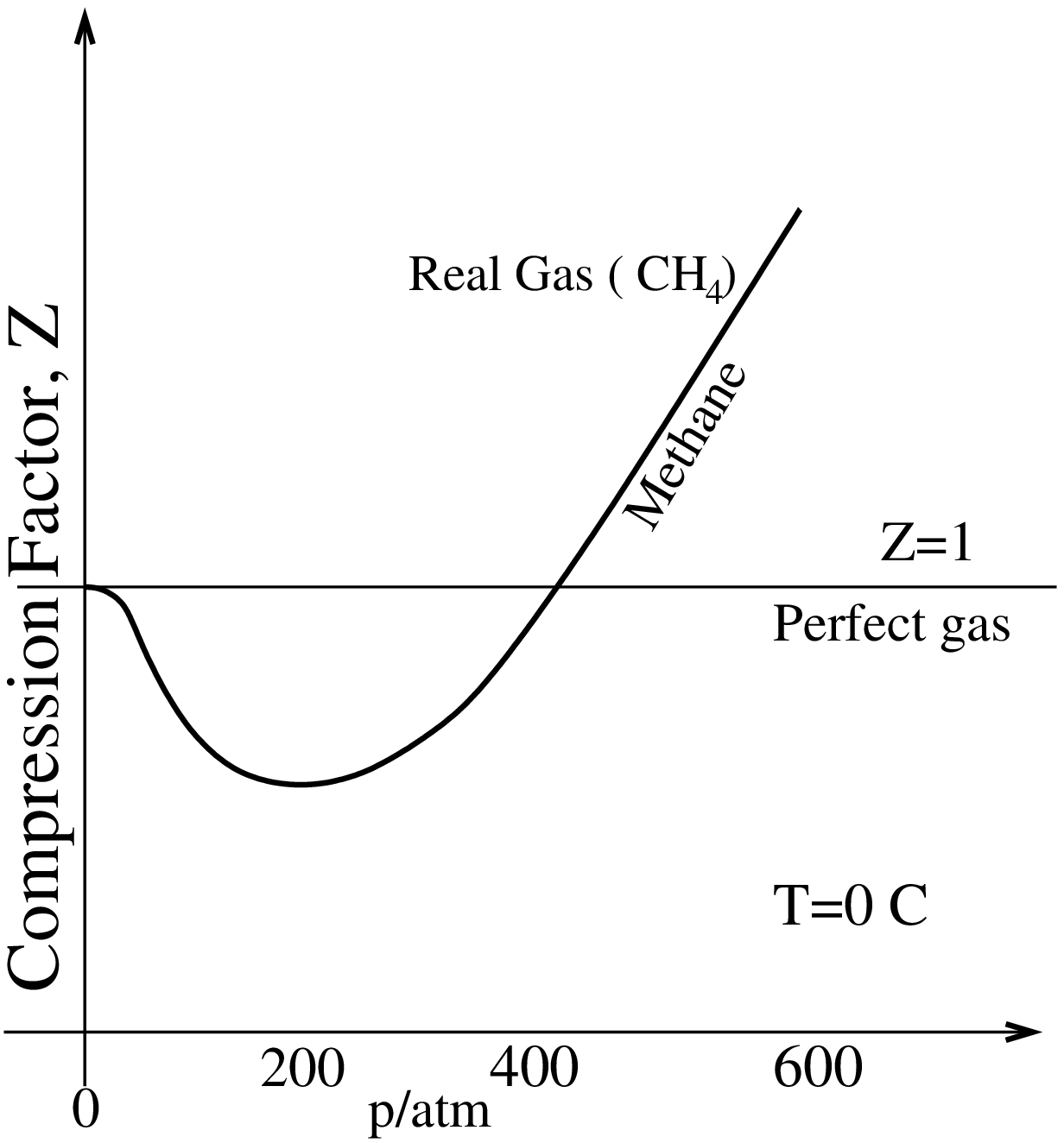}&
\includegraphics[height = 4cm, width =5cm]{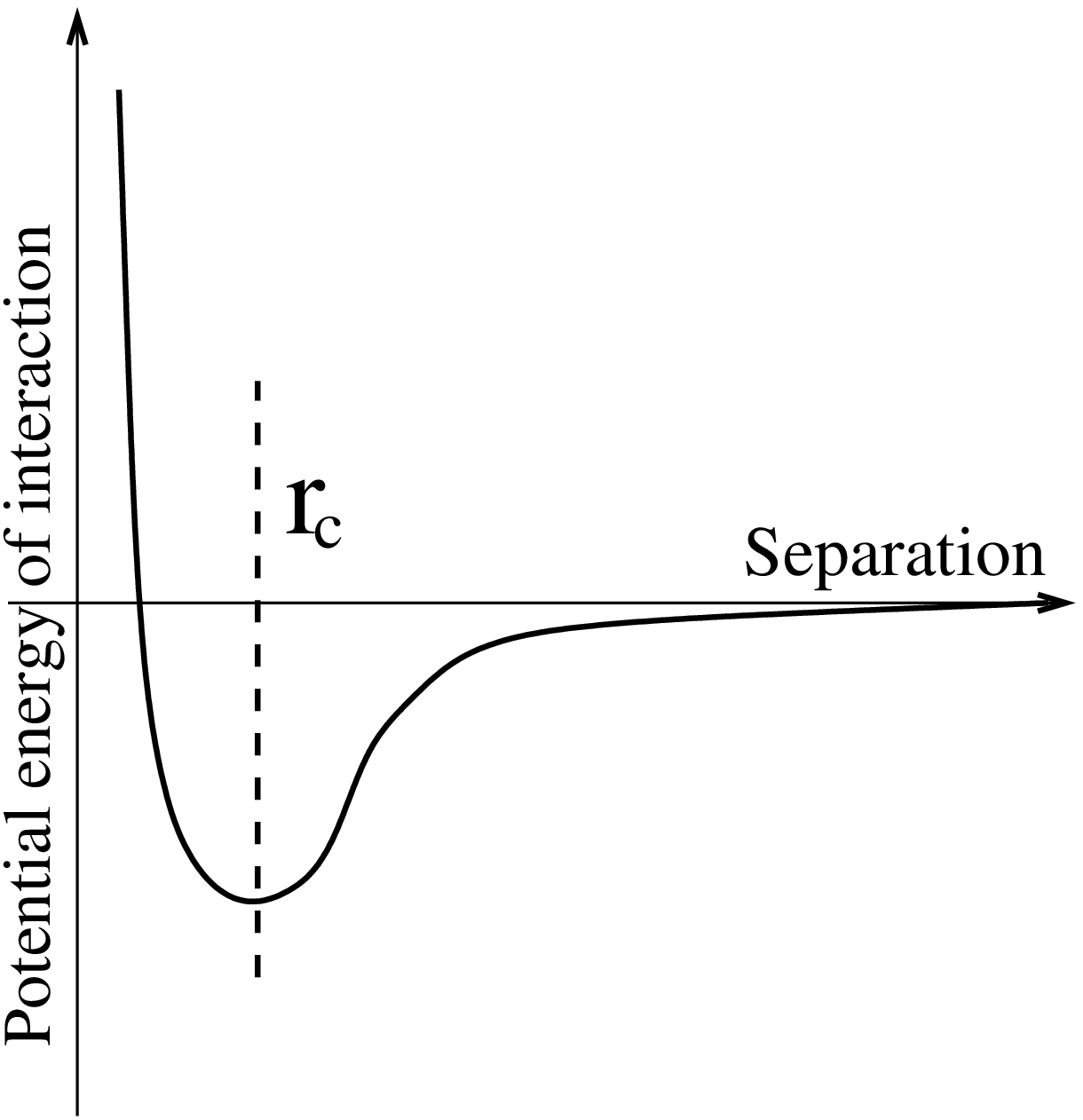}
\end{tabular}
\caption{(a) Behavior of the compression factor for an ideal gas as compared with a real gas.
(b) Inter molecular force as a function of separation in a real gas.}
\label{comp}
\end{figure}

We know that the ideal gas equation
\[PV = N k_B T = n R T\]
shows no phase transition for obvious reasons (due to the neglect of inter-molecular
forces). In contrast, real gases at low temperature and high pressure deviate markedly from
the ideal gas behavior. To see this, let $V_m^0$ be the molar volume (volume per mole) of an ideal gas:
\[p V_m^0 = R T.\]
Define the compression factor $Z$ at a temperature $T$ for a real gas ($P V_m = R T$):
\[Z =\frac{V_m}{V_m^0}= \frac{p V_m}{R T}.\]
In figure \ref{comp}(a) compression factors of both the gases are plotted as a function of pressure.
In an ideal gas, as there is no interaction between the molecules, the compression
factor remains
unity with pressure while for a real gas it first decreases and then at higher pressure it starts
to rise again (figure \ref{comp}(a)). This can be easily correlated with the inter-molecular potential plotted in figure
\ref{comp}(b). At high pressure gas is compressed and molecules are relatively near to each other and thus
exert repulsive forces on each other as the potential is large positive (figure \ref{comp}(b)). This makes the
compression
factor greater than that of an ideal gas. In other words, for the same number of moles of a real gas and an ideal gas in the high pressure limit (when the mean separation between the molecules is less than $r_c$), same pressure will be exerted only if the real gas occupies more volume thus making $Z>1$. In the opposite limit of low pressure (mean inter molecular separation greater than $r_c$), molecules exert
attractive forces and thus smaller molar volume and smaller compression factor for the real gas.
\begin{figure}
\includegraphics[height = 5cm, width =12cm]{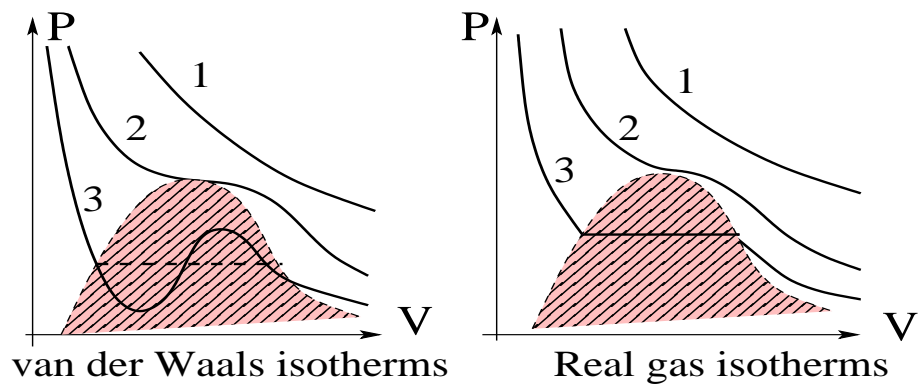}
\caption{Comparing isotherms obtained from van der Waals' equation with those of experiment: curve 1 for $T>T_c$; curve 2 at $T=T_c$; curve 3 at $T<T_c$.}
\label{wloops}
\end{figure}
To account for deviations of the ideal gas law and to understand the phase transition of
gas-to-liquid,  van de Waals postulated a modified equation of state, now called the van der Waals
equation of state\cite{23}. {\it This is a remarkable piece of physical intuition and deduction of an equation of state
from average properties of molecular interactions without going into formal mathematical
development. }This is a physical tour de force and he was awarded with Nobel prize (1910) for this.

\begin{wrapfigure}{r}{0.3\textwidth}
  \vspace{-20pt}
  \begin{center}
    \includegraphics[width=0.3\textwidth]{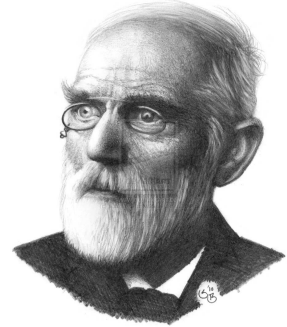}
  \end{center}
  \vspace{-15pt}
\caption{van der Waals (1837--1923)).}
  \vspace{-5pt}
\end{wrapfigure}

He argued
``....as you are aware the two factors which I specified as reasons why a
non-dilute aggregate of moving particles
fails to comply with Boyle's law are firstly the attraction between the
particles, secondly their proper volume...''
----From Nobel prize lecture---JOHANNES D. VAN DER WAALS.

Thus replacing the actual volume and pressure in the ideal gas equation by an effective ones:
\[p_{eff} = \frac{N k_B T}{V_{eff}}\]
\[p = \frac{N k_B T}{V- N b} - a (\frac{N}{V})^2\]
results a mean-field theory. 
This takes ``an average effect'' of molecular interactions, and this type of
theory is called the mean-field theory. As the real molecules repel at short distance, the effective volume is less $V_{eff} = V-N b$ (where $N$ is the number of molecules and $b$ is an effective volume of a single molecule). Similarly effective pressure $P_{eff}$ is less due to long range attarctive interaction between the molecules.
If we compare the isotherms of a real gas with those obtained by van der Waals
equation we get a qualitative agreement. Van der Waals isotherms look qualitatively the same as
those in the experimental one (figure~(\ref{wloops})), but this has two basic defects:
\begin{enumerate}
 \item Unphysical van der Waals loops.
 \item Different critical point behaviour.
\end{enumerate}
The unphysical loops means that if one increases volume pressure increases (for the curve with positive slop in figure ~(\ref{wloops}) in the coexistence regime). This is unphysical and  can be rectified with Maxwell construction\footnote{In Maxwell's construction one replaces the curve in the co-existence regime by a straight horizontal line such that the area under it is equal the area above it (ref. figure ~\ref{wloops} (a))\cite{24}.}. 

\begin{wrapfigure}{l}{30mm}
\includegraphics[height = 3cm, width =3cm]{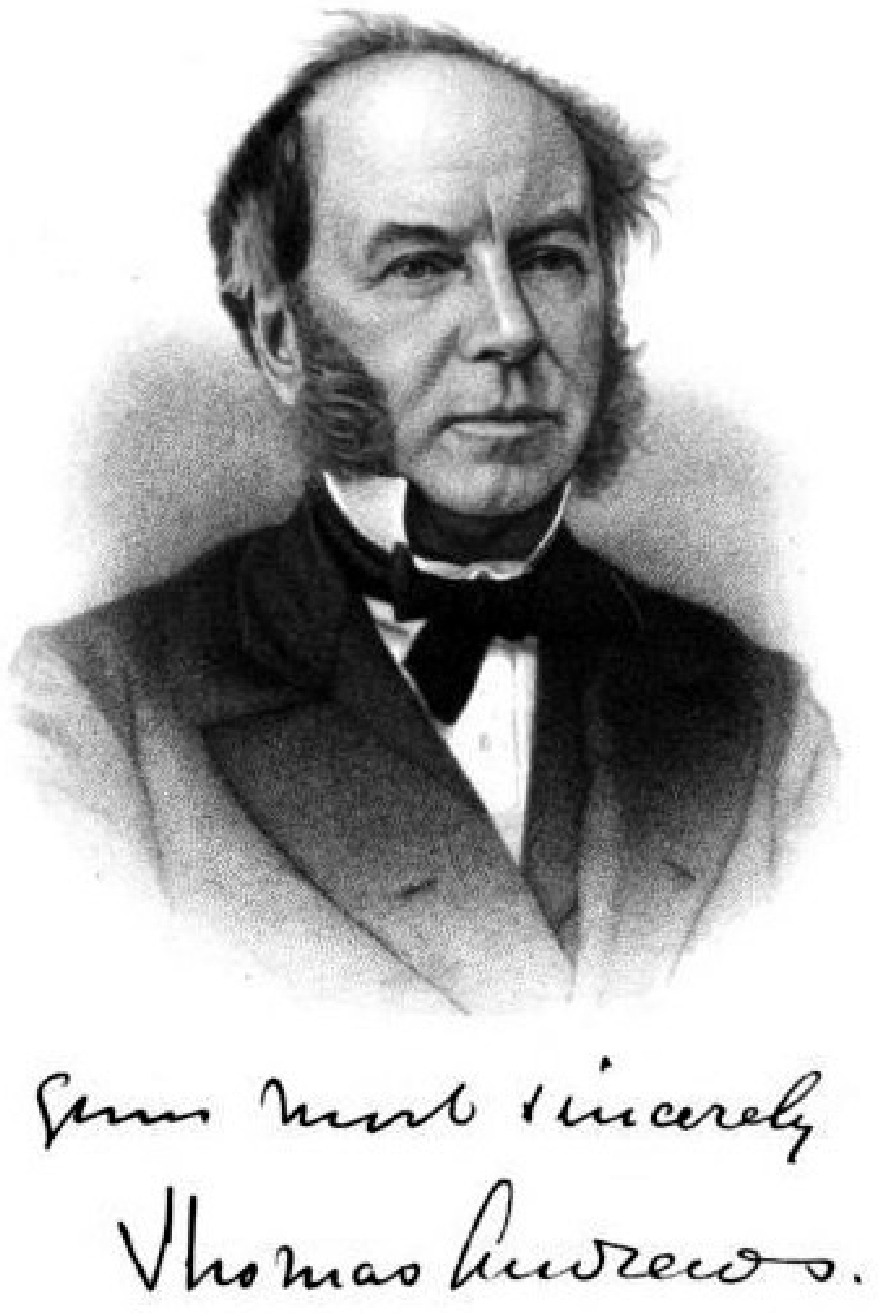}
\caption{Thomas Andrews (1813--1885)}
\end{wrapfigure}

The critical pressure, temperature, and volume can be calculated from:
\[\frac{d p}{dV} = \frac{d^2 p}{d V^2} = 0.\]
And near the critical point $|T_c-T|<< T_c$ and from the above derivatives, it can be easily shown that:
\[\rho_{liquid}- \rho_{gas} \propto (T_c -T)^{\beta}.\]
From van der Waals' theory $\beta$ comes out to be $1/2$ (the mean-field
value). But, in actual experiments on liquid-gas systems, $\beta = 1/3$. This was first
experimentally deduced by T. Andrews  in 1869! This very clearly showed the contradiction of the
mean-field theory with experiment. 

We will see that this discrepancy is not specific to van der Waals theory but a basic defect of all such theories called the MFTs at the critical point. It took about $100$ years for the resolution of this discrepancy!! This happened in the hands of K.
G. Wilson in early $70$s. The culprit is non-negligible fluctuations (density fluctuations in this
liquid-gas case) at all length scales that invalidates the application of the mean-field theory
itself\cite{6}.

The second example of MFT we consider is that of the famous Ising model----a caricature for spontaneous ferromagnetism (like in Iron).
\begin{figure}[h!]
\centering
\begin{tabular}{cc}
\includegraphics[height = 4cm, width =6cm]{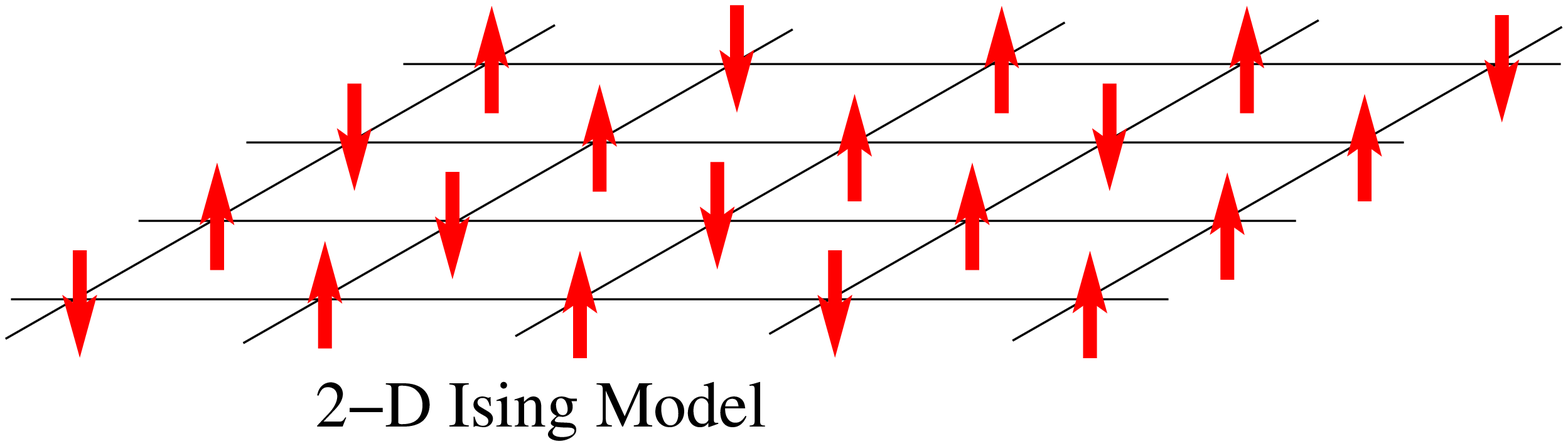}&
\includegraphics[height = 5cm, width =5.5cm]{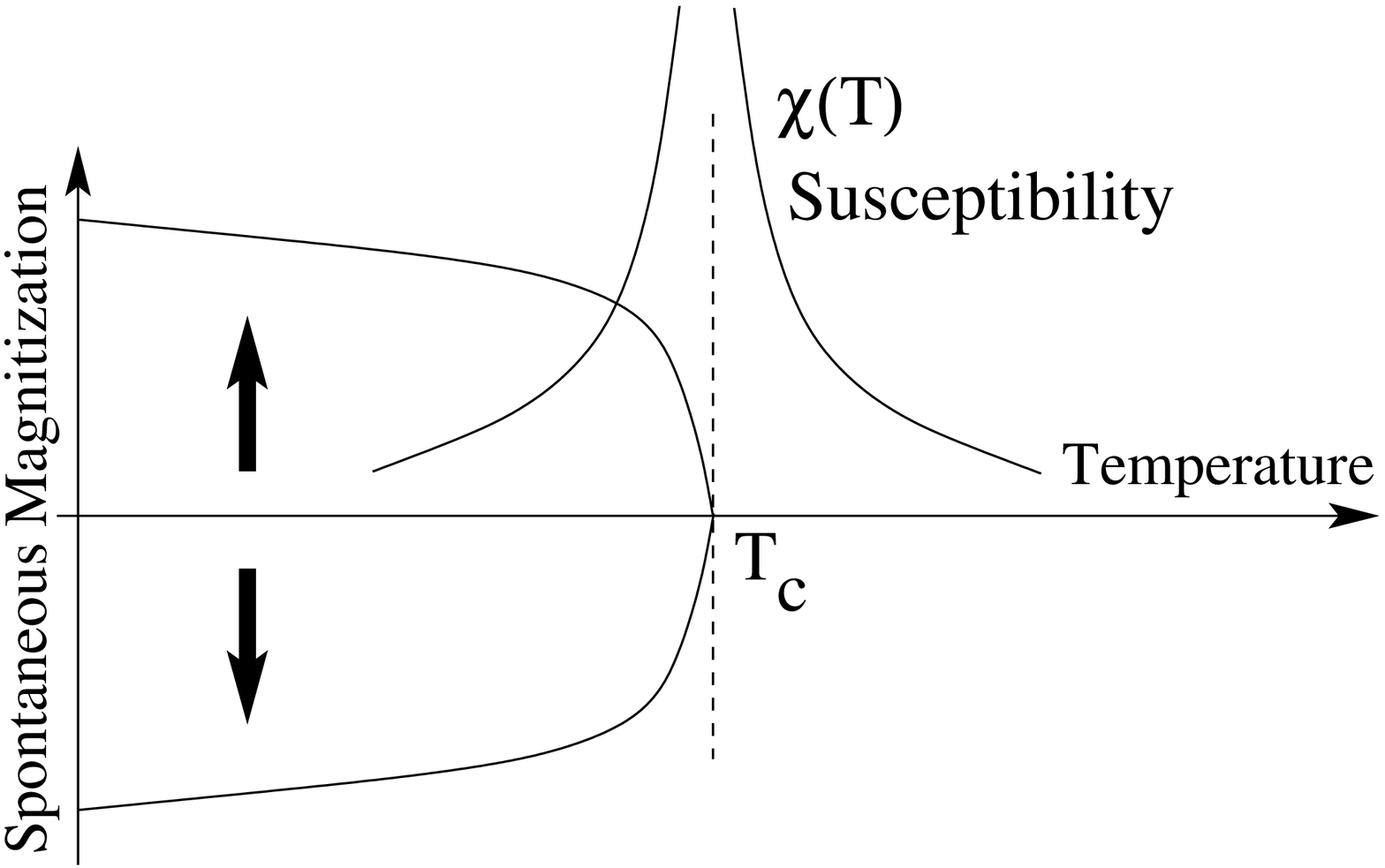}
\end{tabular}
\caption{(a) 2-D Ising model (b) Phase diagram of the model.}
\label{ising}
\end{figure}
A simple two dimensional Ising model is depicted in figure ~\ref{ising}(a) in which spins are represented by
arrows and these interact with their nearest neighbors. Each spin can have two states, either up, or down. This short range interacting model shows the
phenomenon of spontaneous magnetization. Consider that the model is immersed in an external magnetic field $H$. Write the Hamiltonian in dimensionless form:
\[\frac{H}{k_B T} = - K \sum_{<nn>} \sigma_i \sigma_J - h \sum_i \sigma_i\]
Here $K =\frac{J}{k_B T}$ and $h= \frac{\mu H}{k_B T}$. There are two competing tendencies:
\begin{enumerate}
 \item Lining-up of spins due to mutual spin-spin interaction ($K$) and the external field ($h$).
 \item Disruption due to thermal agitation.
\end{enumerate}
At low temperatures first tendency wins over and at high temperature the second one wins over. To calculate mean magnetization (a measure of the lining-up of the spins), one  computes the Free Energy in the standard way:
 \[F = - k_B T \ln\left(\sum_{[all~configurations:~ c]} e^{- H(c)/k_B
T}\right).\]
Here comes the big question: Why do statistical mechanical systems show phase transitions
(non-analytical behavior like spontaneous magnetization, see figure \ref{ising}(b); the graph near $T=T_c$) when the system's free
energy is a smooth function of temperature and of system's parameters (an
analytical function)? Sum of smooth exponentials is a smooth function!

The origin of non-analyticity (like spontaneous magnetization) is in the
thermodynamical limit (infinite system size!): Larger the system size sharper is the transition:
As real phase transitions involve thermodynamical systems having very large number of
particles of the order of Avogadro  number, {\it transitions appear infinitely sharp to human
perception!}  This has been dubbed as ``extended singularity theorem'' by Leo
Kadanoff or better one can say ``extended singularity conjecture''\cite{3}. As the proof exists only in one special case\cite{25}.

Mean-field theory for the Ising model is done in usual way\cite{3,26,27}:
\begin{enumerate}
\item First, consider that just one spin is immersed in magnetic field  (that can take two values:
$+1$ (up-spin, say), and $-1$
(down-spin)). Thermal  expectation value of the spin is:
 \[\la \sigma \ra = \frac{\sum_{\sigma = \pm 1} \sigma e^{h\sigma}}{\sum_{\sigma
= \pm 1} e^{h \sigma} } = \tanh(h).\]
\item Second, when our test spin is interacting with many neighboring spins,
all immersed in magnetic field:
\[\la \sigma \ra = \frac{\sum_{\sigma = \pm 1} \sigma e^{h_{eff} \sigma}}{\sum_{\sigma
= \pm 1} e^{ h_{eff} \sigma}} = \tanh(h + K \sum_{<nn>}\sigma_i) \simeq \tanh(h_{eff}).\]

\[h_{eff} = h + {\rm effective~field~due~to~other~spins}= h+ z  K ~\la \sigma \ra .\]
Here the exact field $K \sum_{<nn>} \sigma_i$ due to near-neighbor spins is
replaced by the effective field $\sim z K \la \sigma \ra$ ($z$ is called the
coordination number, $=4$ in 2D). This constitutes the mean-field
approximation.
\end{enumerate}

\begin{figure}
\includegraphics[height = 3cm, width =10cm]{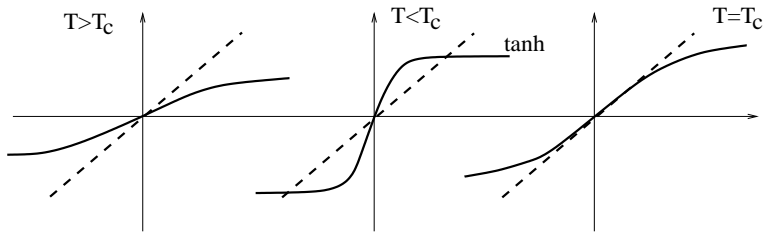}
\caption{Solutions of the zero field mean-field equation.}
\label{soln}
\end{figure}

Thus in MFT: $\la \sigma \ra = \tanh(h + z K \la \sigma \ra)$. To see what values $\la \sigma\ra$ can take, one must solve this self-consistently. As
$K\propto\frac{1}{T}$ it turns out that $z K =\frac{T_c}{T}$ (where $T_c$ is the critical
temperature). In zero external field $h=0$ the equation $\la \sigma \ra = \tanh(\frac{T_c}{T} \la \sigma \ra)$
has the following solutions (figure~\ref{soln}): 
\[ 
\la\sigma\ra = \left\{
\begin{array}{l l}
0          &  if  ~~T>T_c\\
0, +1, -1  &  if ~~ T<T_c\\
degenerate &  if  ~~T=T_c\\
\end{array}\right
.\]
The region $T\leq T_c$ is of particular importance.  Near the critical point $\la\sigma\ra\simeq 0$, $\tanh(x)$ can be expanded for small $x$ as $\simeq x-\frac{1}{3}x^3+...$. And it can be easily shown that near the critical point:
\[\la\sigma\ra \propto (T_c-T)^\beta.\]
With $\beta =\frac{1}{2}$ (the mean field value). Compare this with van der Waals theory! (THE VERY
SAME BETA VALUE!)

Experiments on a wide variety of magnetic systems show that $\beta \simeq \frac{1}{3}$. Again, we
see the contradiction between mean-field theory and experiments. So, something is wrong with
mean-field theory near the critical point! But, as mentioned before, solution comes much later. However, the great merit of MFTs is that they qualitatively explain the emergence of phases below $T_c$ (refer to the middle graph in figure~(\ref{soln})).  One can show that the free energy for $\la \sigma \ra = 0$ phase is greater than the other phases with non-zero $\la\sigma\ra$, and the phase with minimum free energy is the phase that is realized in a thermodynamical system. Thus one has non-zero $\la\sigma\ra$ for temperature less than the critical temperature (figure \ref{ising}).

\section{Landau's paradigm: symmetry breaking and the order parameter
concept}
\begin{wrapfigure}{l}{40mm}
\includegraphics[height = 4cm, width =4cm]{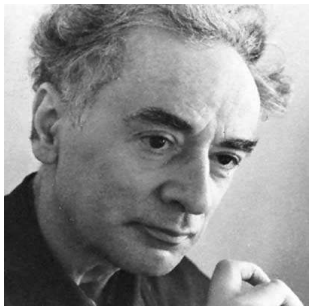}
\caption{Lev. Landau (1908--1968)}
\end{wrapfigure}
In a landmark paper\cite{1} of 1937 Lev Landau formulated the phase transition problems from the
broken symmetry perspective.  {\it All the above phase transition theories can be written  in one mathematical language near the critical point.} He noticed that most of the phase transitions involves breaking of
discrete or continuous symmetry and introduced a very important concept to capture the nature
and extent of this symmetry breaking: the order parameter concept. One can appreciate this rather
formal language  related to symmetry breaking by considering some concrete examples. Before we do that, let us first consider two
important theorems related to symmetry and its breaking. Emy Noether in 1918 put forward an
important theorem: The Noether theorem (1918):

\begin{figure}
\begin{tabular}{cc}
\includegraphics[height = 4cm, width =4.5cm]{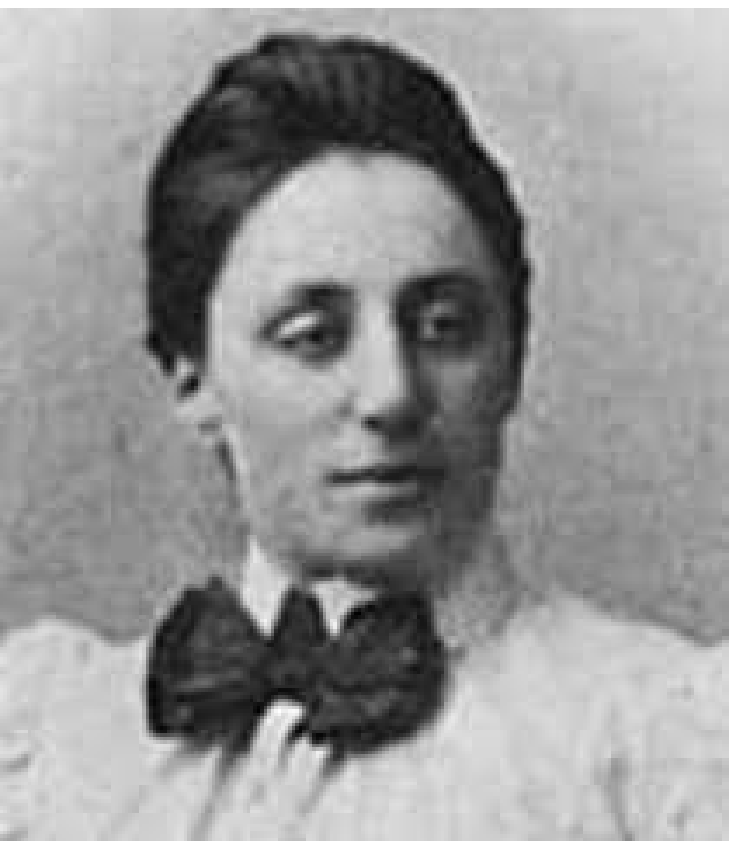}&
\includegraphics[height = 4cm, width =4.5cm]{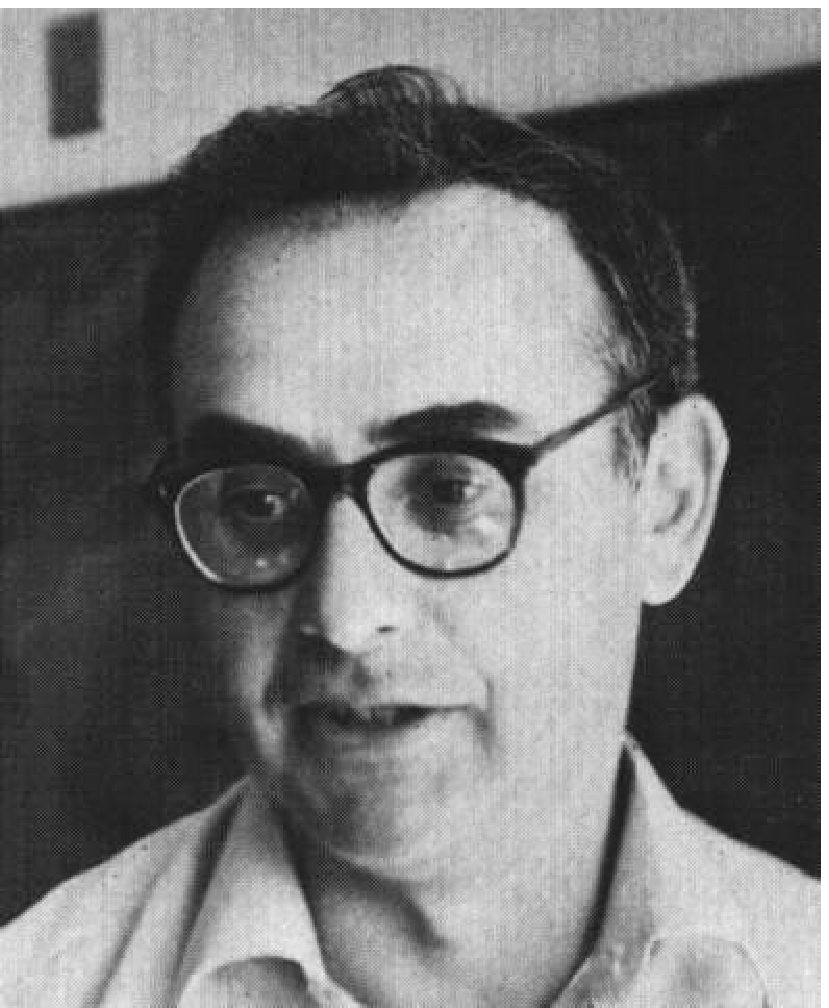}
\end{tabular}
\caption{Emmy Noether (1882--1935) and Jeffrey Goldstone (born 1933).}
\end{figure}

``Every continuous symmetry of a system can be associated with a conserved quantity.'' This theorem
is the cornstone of classical mechanics. There are numerous example, consider for example; if,
gravitational potential along x-direction is constant, then linear momentum along that direction is
conserved and so on. Another important theorem, for the present context, is the Goldstone
theorem (1961):
``The spontaneous breaking of a continuous symmetry can be associated with a spinless and
massless particle--the Nambu-Goldstone boson (or simply the Goldstone Boson)''. To illustrate this we consider few examples:

Example No. 1:
\begin{enumerate}
 \item liquid-to-Crystal transition: In the liquid state atoms/molecules are
under the effect of mutual potential (like van der Waals potential) which does not break the
translational and rotational symmetry of the space (molecules/atoms are free to wander around and rotate).
\item In the crystalline state, a periodic potential emerges, which break the continuous symmetry
under translations (however, this leads to a discrete translational symmetry due to the
periodicity of the lattice: the crystal momentum)\cite{17}.
\item Since in the above example {\it continuous} translational symmetry is broken, momentum
 does not remain a conserved quantity. And this violation of the conservation of momentum in the crystalline state leads a Goldstone boson called the phonon (lattice vibrations).
\end{enumerate}
Example no. 2:
Ferromagnetic transition and broken rotational symmetry: In a ferromagnetic system above the transition temperature, spins are randomly oriented in all possible directions. But below the transition temperature, this rotational symmetry is broken and spins align in the same direction. The Goldstone boson in the ferromagnetic case is the spin-wave excitations (or, collective excitation of spin waves).

One can consider many more examples. Thus each phase transition is a manifestation of some kind of broken symmetry:
In ferromagnetic rotational symmetry is broken; in crystal formation translational symmetry is
broken; in superconductor or superfluid electromagnetic gauge symmetry is broken....
On this basis, Landau  introduced the very important concept of an order parameter.
For example, in ferromagnetic transition the order parameter is an average magnetization per
unit volume (which is zero for $T>T_c$ and non-zero for $T<T_c$); in liquid-gas transition it is the density difference; in superconductor it is the
expectation value of the Cooper field $<\psi_\sigma^\dagger \psi_{-\sigma}^\dagger>$; and in
electroweak symmetry breaking it is expectation value of Higgs field $<\phi>$.... Order parameter is zero in symmetrical phase and is non-zero in the unsymmetrical phase. It is a macroscopic variable, much like a hydrodynamical variable. It has discontinuous jump for first order transitions as the order parameters are the first order derivatives of the free energy (for example, $M = -\frac{\pr G}{\pr H}$). Thus one can connect Landau's scheme with that of Ehrenfests' and one can interpret Ehrenfests' scheme in terms of continuity or discontinuity of order parameters. 
For the second order transitions order parameter evolve from zero in symmetrical phase to
non-zero in the unsymmetrical phase in continuous manner (thus 2nd order transitions are also
called continuous transitions).  Landau's great contribution in unifying the theoretical understanding of wide variety of phase
transitions lies in his series expansion of free energy as a function of order parameter near
the critical point where the order parameter is small.
\[G(M,T) = G_0(T) + r(T) M^2 + u(T) M^4 + ...\]
Where $M$ is some order parameter (like magnetization in ferromagnetic case). The fundamental
assumption that goes into this is the analyticity of free energy as a function of order parameter
(magnetization here) and temperature. This fundamental assumption gets its justification
a-posterior! Most importantly, this is in accord with "extended singularity conjecture" in which free energy per unit volume remains finite in the thermodynamical limit ($N\rta \infty,~V\rta \infty,~~\frac{N}{V}=finite$). The sharpness of phase transitions (apparent non-analyticity) is something to do with very large number, of the order of Avogedro number, of particles present in the system.

Landau argued, according to the basic principle of thermodynamics, that at a given temperature, the phase with minimum free energy is the stable one: $\frac{\pr G}{\pr M} = 0$  leads to:
\[r(T) M + 2 u(T) M^3=0.\]
Which implies that either $M=0$ or $M = \pm \sqrt{-\frac{r(T)}{2 u(T)}}$.

\begin{wrapfigure}{r}{30mm}
\includegraphics[height = 5cm, width =4cm]{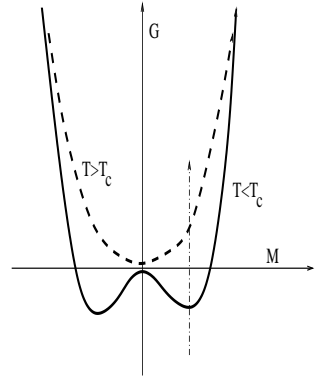}
\caption{Behavior of free energy in Landau's formulation.}
\label{landauf}
\end{wrapfigure}

One can relate these to the following two possible cases:
\begin{enumerate}
\item Case A: If $r(T)>0$: one has only one minimum at $M=0$ ($u(T)$ has to be positive as free
energy is bounded from below). $M=0$ in this case and it corresponds to symmetrical phase ($T>T_c$).
\item Case B: If $r(T)<0$: one has two minima at  $M = \pm \sqrt{-\frac{r(T)}{2 u(T)}}$ and this
corresponds to the unsymmetrical phase ($T<T_c$) (see figure ~(\ref{landauf})) with non-zero value of the order parameter.
\end{enumerate}
From above, one can deduce the critical exponent $\beta$ as follows:
\begin{enumerate}
\item As $r(T)$ is positive for $T>T_c$ and negative for $T<T_c$, it must pass through zero
at $T=T_c$.
\item So near the critical temperature $r(T)$ can be expanded as: $r(t) = a (T-T_c) + O((T-T_c)^2)$.
\item Thus magnetization  in $T<T_c$ phase can be written as: $M =\pm \sqrt{\frac{a(T-T_c)}{2 u(T)}} \sim
(T-T_c)^\beta$. Here $\beta = \frac{1}{2}$.
\end{enumerate}
Again $\beta$ takes the same value $1/2$, as in van der Waals theory and in the mean-field
theory of the Ising model. The general disagreement of MFTs with experiment is clear. We will see in the next section that systems near criticality violates an important property of statistical independence and the correlation length diverges at the critical point. Before we embark upon the question of the behaviour of correlation length and renormalization group we first would like to consider the question of the critical point in the next section in the light of  some exact solutions.

\section{Problems of the Mean-Field Theories and Landau's
paradigm}

In the previous section we have seen the contradiction of MFT with experiment near a critical point. The $\beta_{theory}=\frac{1}{2}$ and $\beta_{experiment}=\frac{1}{3}$. This was apparent even in the first decade of 20th century.  Still, MFTs
were developed (like Curi-Weiss, Ising model, Landau's generalization) and trusted.
Nobody attempted to resolve the problems of MFTs at the critical point\cite{3}. Then in 1944,  with the Onsager's---a mathematical tour de force---{\it exact} solution of the 2-D
Ising model, it became very clear that something is very much wrong with MFTs near criticallity:

\begin{enumerate}
\item C. N. Yang with an extension of Onsager's solution found that $\beta$ is $1/8$ different
from mean-field result $\beta=1/2$.
\item  Onsager's solution showed that heat capacity diverges as a logarithm of $T-T_c$ whereas MFT
shows a finite jump in heat capacity at $T_c$.
\item The correlation length $\xi$ diverges as $\frac{1}{\sqrt{T-T_c}}$ in MFT whereas it diverges
as $\frac{1}{T-T_c}$ in the Onsager's solution. Thus, $\nu$, the critical index of the correlation length, is one half in MFT whereas it is equal to one in the Onsager's solution!
\end{enumerate}

Here comes the 2nd big question: why do MFTs fail to predict the correct values of critical indices? The answer lies in the fact that mean-field equation $\la\sigma\ra = \tanh(h + \frac{T_c}{T}\la\sigma\ra)$ fails to take into account the longer length scale fluctuations (as compared to inter-particle and lattice constant). See for details\cite{6}.

As fluctuations at all length scales (from lattice constant to the correlation length) become important near a critical point, a proper theory in this regime must address this issue. The theory is called the renormalization group (next section).

\section{Going beyond Landau's paradigm--The renormalization (semi)group (RG)}

The renormalization theory in general deals with problems of long correlation length. The presence of long correlation length leads to the breakdown of an important property of statistical mechanical systems called Statistical Independence (SI). Thus the results of standard Gibbs-Boltzmann statistical mechanics become unreliable. Near the critical point correlation length diverges as can be easily seen: 

Consider that the order parameter is space dependent $M(x)$ and vary very slowly as compared to
atomic dimensions: One writes Ginzburg-Landau free energy functional\cite{2}:
\[F = \int d^3 x \{ (\nabla M(x))^2 + r(T) M(x)^2 + u(T) M(x)^4 - B(x) M(x)\}.\]
Correlation length $\xi$ is defined as the distance over which the effect of a test spin extends
out to other nearby spins. To calculate $\xi$ consider $B(x) = B_0 \delta^3(x)$ (magnetic field in a narrow region of space).
Neglect the smaller 4th order term and by minimizing $F$ with a variational calculation, one
 readily obtains:
\[-\nabla^2 M(x) + r(T) M(x) = B_0 \delta^3(x).\]
With Fourier transforms one gets an immediate solution:$ M(x) \sim \frac{e^{\sqrt{r(T)} |x|}}{|x|}.$
From which one defines a correlation length $\xi = \frac{1}{r(T)} \propto
\frac{1}{(T-T_c)^{1/2}}$. Thus correlation length diverges at the critical point with critical index $\nu = 1/2$.

Due to this diverging correlation length various parts of the system respond in a correlated way. This makes the system non-self-averaging! The fundamental hypothesis (actually an important property of ordinary statistical mechanical systems with short range interactions) of statistical independence as highlighted by Landau\cite{5} cannot be applied and fluctuations in the additive observables (sum-functions) do not obey $\frac{1}{\sqrt{N}}$ law\cite{5}. In other words, fluctuations could be significant as compared to the average behavior. A comparison of physical problems with short and long correlation length is presented in table~(\ref{phys})
\begin{table}[h!]
\caption{Physical problems with short and long correlation length}
\begin{tabular}{|p{5.5cm}||p{5.5cm}|}
\hline
\multicolumn{2}{|c|}{Physical Problems} \\
\hline\hline
Problems with short correlation length ($\xi<<L$)& Problems with large correlation length ($\xi\sim
L$) \\
\hline\hline
Various parts of the system with size $<\xi$ behaves statistically independently & As
correlations extend throughout the system, various parts of it are statistically correlated! \\
\hline
Statistical independence (SI) leads to product distribution functions:
$\rho_{12} = \rho_1\rho_2$& $\rho_{12} \ne \rho_1 \rho_2$ \\
\hline
Entropy and Energy of sub-parts become additive (weakly interacting sub-parts of size $\xi$)
&non-additivity of entropy and energy of sub-parts\\
\hline
Standard statistical mechanics can be applied &Results of standard statistical
mechanics are not reliable!\\
\hline
Fluctuations in observable converge $\sim \frac{1}{\sqrt{N}}$ & As SI breaks down,
fluctuations in observable diverge!\\
\hline
Examples: systems away from criticallity: MFTs; Hartree-Fock; Cluster and virial exp...&
Thermodynamical systems near criticallity; QCD; Knodo problem; cuprate
superconductivity etc.\\
\hline
\end{tabular}
\label{phys}
\end{table}
The problems with large correlation length are  among the hardest problems in theoretical
physics. For example, thermodynamical systems near criticallity; Kondo problem; high temperature
superconductivity; relativistic quantum field theory; strong interactions and QCD etc.
And most importantly, systems with large correlation length show a certain kind of cooperative
behavior in which the details of the microscopic interactions play secondary role! Before we explore the renormalization group as formulated by Wilson, we first review the essence of a key work due to Kadanoff\cite{28,3} on which Wilson highly relied.

\subsection{Kadanoff's great insight}

\begin{enumerate}
\item Consider for  concreteness the 2-D Ising model. Away from critical point ($T=T_c$) the
correlation length is small ($\xi \propto \frac{1}{(T-T_c)^\nu}$), fluctuations on a smaller length scale are important and one can reliably use the mean field theory.
\item At  $T=T_c$ the correlation diverges, as shown before. A reliable picture can only be obtained if one take all
the spins into account in an area $\sim \xi^2$ (A HOPELESS TASK !).
\item Kadanoff approached the problem with a brilliant idea: map the problem of large correlation
length to a problem with small correlation length!!!!
\item This is possible by exploiting the fact that on smaller lengths scales $l<<\xi$
spins are well
correlated!
\end{enumerate}
To execute the point number 3 above, consider the original lattice (figure~\ref{blocks}) with lattice constant $=a$ in which we mentally construct blocks, each containing four spins (dashed squares in the figure). We then assign an effective spin to each block--the block spin.  The effective spin is an average of the original spins (one can use simple majority rule, if majority is up the effective spin or block is also up). The ease with which we can define block spins is granted by the point 4 above.  If all the spins are up in a given block, then our effective spin is four times the  original spin. The  resulting lattice of this effective block spins will have lattice constant $a_1 = 2 a$. The block spins or effective spins will interact with each other with some modified interaction that one has to calculate from the detailed model. The correlation length scales as follows:
\begin{figure}
\includegraphics[height=6cm,width=10cm]{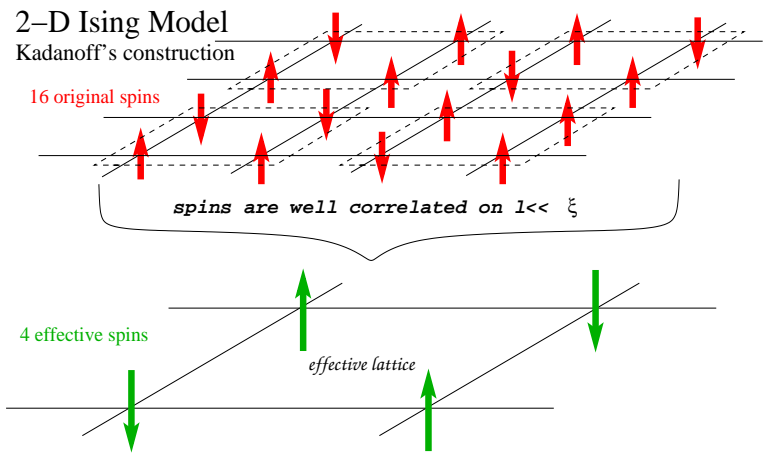}
\caption{Kadanoff's Block spin renormalization idea}
\label{blocks}
\end{figure}
If the $\xi$ is $1000~a$ (in the original lattice), then it will be $500 ~a_1$ (in the scaled
lattice).

\begin{figure}
\centering
\begin{tabular}{cc}
\includegraphics[height = 4cm, width =5cm]{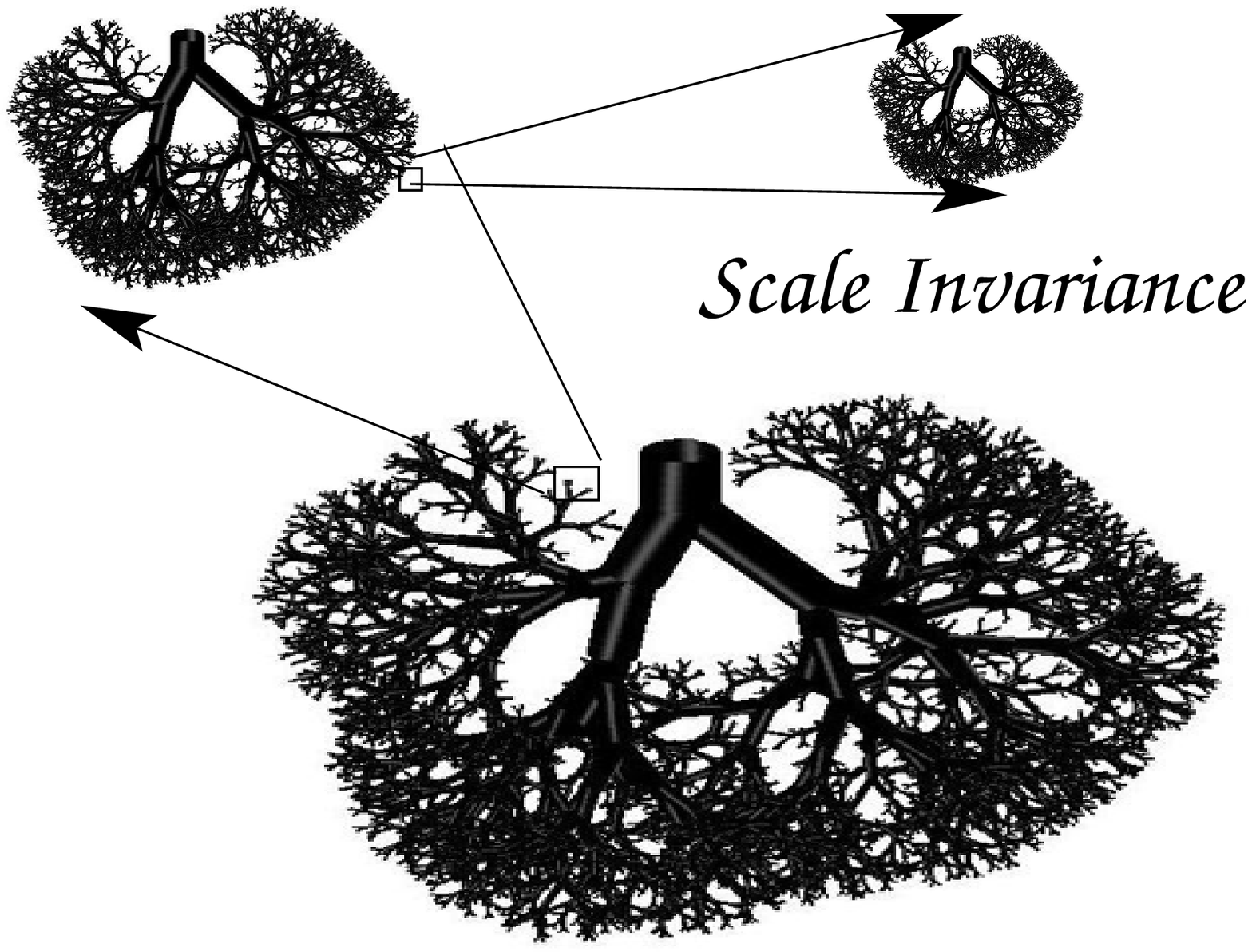}
\includegraphics[height = 4cm, width =5cm]{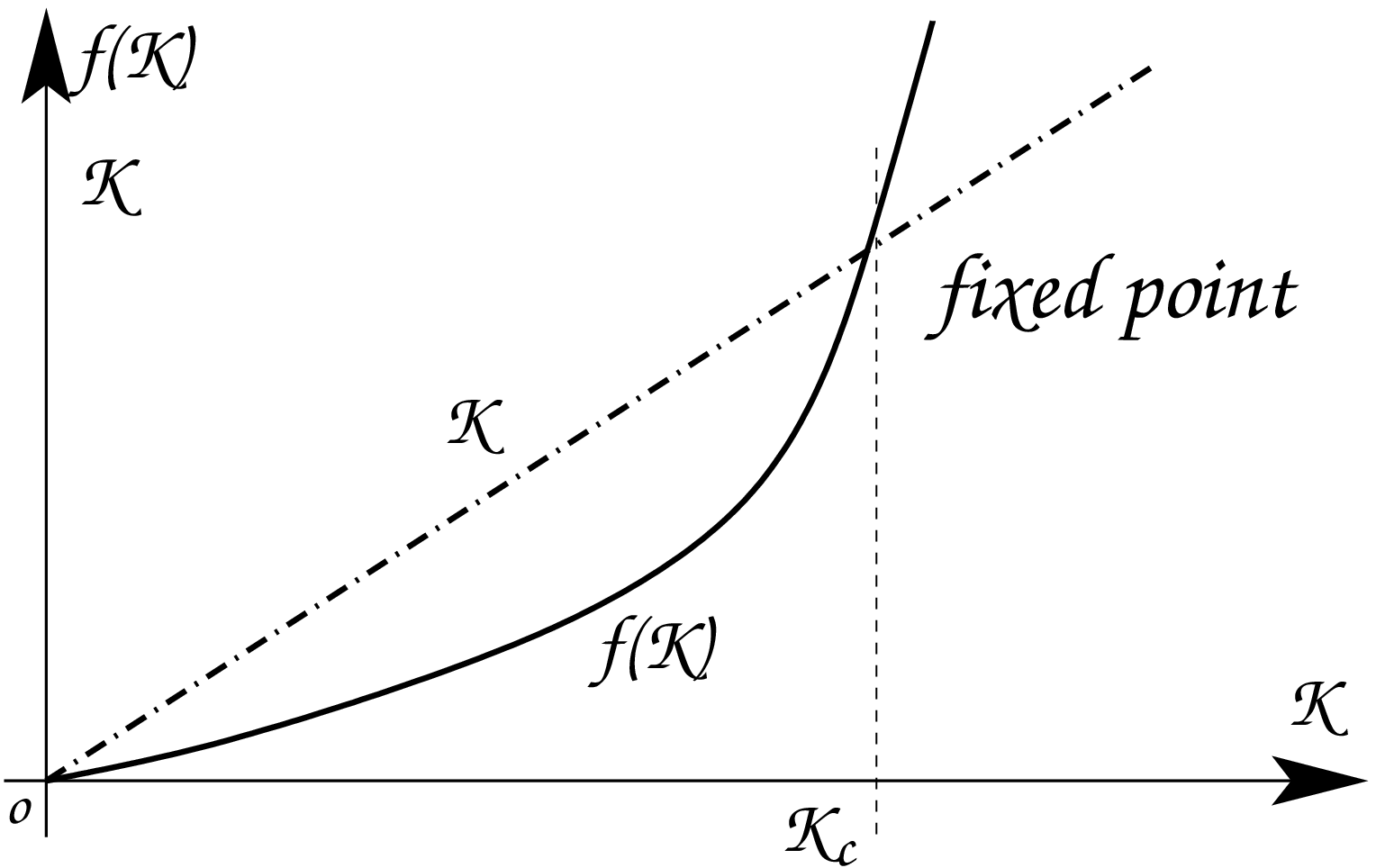}
\end{tabular}
\caption{(a) Scale invariance: same kind of structure re-appears at all length scales when correlation length diverges (if one magnifies any part of the system, then it is not possible to notice the difference between the magnified part and the original system). (b) Fixed points: self-consistent solution of $K=f(K)$.}
\label{scale}
\end{figure}

Thus the problem with $\xi=1000$ is reduced to the problem with $\xi =500$! Repetition of this procedure ultimately leads to $\xi\sim 1$ (for a finite system). But for a system in thermodynamical limit (system of infinite extent) one can exploit the following
fundamental fact: ``$\xi$ diverges at the critical point'' This fundamental fact in an infinite system leads to a very curious behavior of system's couplings. They tend towards a fixed point as explained below:

Let the Hamiltonian of un-scaled system be $H = \sum_{n,i} K_0 s_n s_{n+i}$. Here $K_0$ is near neighbor interaction in the un-scaled lattice.  After first scaling, let $K_0\rta K_1^{eff}$ and $s_n\rta s_n^{(1)}$ (i.e., block spins
$s_{n}^{(1)}$ interact with effective coupling $K_1^{eff}$). Thus after first scaling:
\[ \sum_{n,i} K_0 s_n s_{n+i} \rta \sum_{n,i} K_1^{eff} s_n^{(1)} s_{n+i}^{(1)}.\]
After $m$ repetitions:
\[H^{(m)}_{effective} = \sum_{n,i} K_m^{eff} s_n^{(m)} s_{n+i}^{(m)}.\]
The correlation length behaves as follows:

After first scaling $\xi^{(1)}(K_1^{eff}) = \frac{1}{2} \xi(K_0)$ (If the $\xi$ is $1000~a$ (in the original lattice), then it will be $500 ~a_1$ (in the scaled
lattice)). After $m$-scalings : $\xi^{(m)}(K_m^{eff}) = \frac{1}{2^m} \xi(K_0)$. Now the fundamental assumption that goes in Kadanoff treatment is this: {\it The scaling does not
change the physics of the problem}, namely, $\xi^{(m)}(K_m^{eff})$ diverges as $K_m^{eff}\rta K_c$ as the original correlation length
$\xi(K_0)$ does at $K_c$ ($K_c = \frac{J}{k_B T_c}$). If one knows the functional form $K_{m}^{eff} = f(K_{m-1}^{eff})$ for one transformation, the same functional form will be true for all scalings!---precisely at the critical point $K_0=K_c$ (where correlation length is infinite). This scale invariance (figure~\ref{scale}(a)) leads to $m$ independent relation (AT THE CRITICAL PONIT):
\[K_c =f(K_c.)\]
The self consistent solution to this equation leads to the critical point solution (or fixed point
solution) that gives $T_c$ (figure~\ref{scale}(b)). Thus from the Kadanoff picture $T_c$ can be calculated if one knows the function $f(K)$. More importantly, the critical index $\nu$ can be expressed in terms of the derivative of $f(K)$ wrt $K$ at $K=K_c$: Suppose $K$ is near to $K_c$, then $f(K) = f(K_c) + (K-K_c) \lambda$. Here $\lambda = \frac{df}{dK}$. As $K_c =f(K_c)$, implies $f(K)-K_c = \lambda (K-K_c)$. Now suppose correlation length behaves as $\xi(K)\sim (K-K_c)^{-\nu}$, then one must have\[\frac{\xi(f(K))}{\xi(K)}=\left(\frac{f(K)-K_c}{K-K_c}\right)^{-\nu}\]
From the Taylor expansion above, it immediately follows that $\frac{1}{2} = (\lambda)^{-\nu}$ as the scaling factor is 2. Thus if $\lambda$ is known then $\nu$ can be determined (see for other complications\cite{2}).

The problem of the Kadanoff's procedure is that it does not give the central method to determine the important  function $f(K)$. He used the above argument to derive scaling relations\cite{28}. Important point worth noting is that Kadanoff approach is radically different from traditional statistical mechanical approach of constructing ensembles and using probability distribution functions.

This qualitative picture of Kadanoff's was turned into an efficient computational procedure with
far reaching further physical insights (origin of universality etc) by Ken Wilson in early $70$s. This approach is known as the renormalization group approach of Ken Wilson\cite{2}. Although the
concept in its primitive form and with certain drawbacks was know in the early days of quantum field
theory (QFT) for example, Gell-Mann and Low etc,   it was new approach in critical phenomena (in 1970s) that
resolved the problems of MFTs in describing critical indices. We will clarify the Wilson's approach using one specific example, the $\phi^4-$theory,
although the approach has much more general applicability. The $\phi^4$ theory also connects to our
previous example of the  Ising model!

Next subsection is devoted to Wilson's formulation. A brief biographical sketch of the man behind this great feat is given in the appendix.

\subsection{Wilson's formulation}

As promised in the last subsection we would like to answer the following question:
{\textsc{How did  Wilson change Kadanoff's qualitative scheme into a
powerful computational method?}} For this we consider the $\phi^4$- theory.

\begin{figure}
\includegraphics[height = 3cm, width = 7cm]{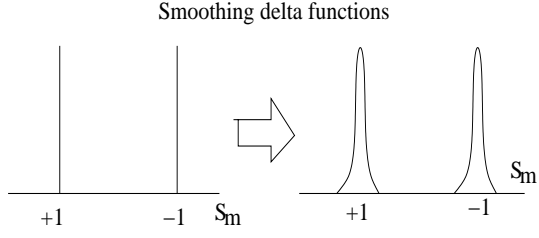}
\caption{Smoothing the delta function peaks into narrow Gaussian peaks.}
\label{delta}
\end{figure}

\underline{The $\phi^4--$ Theory:}
Consider again the example of the Ising model (figure~\ref{ising}) in zero external magnetic field:
\[\frac{H}{k_B T} =  -K \sum_{<nn>} \sigma_i\sigma_j.\]
We stress that if one wants to understand the physics of the renormalization group one must work through the mathematical apparatus which is rather standard now; namely, the functional integrals and the method of contractions. Below the logic of the calculation and essential steps are given (see for details the original reference; Wilson-Kogut in\cite{2}). The partition function $Z$ for the Ising model can be written as:
\[\prod_m \int_{-\infty}^{+\infty} ds_m 2 \delta(s_m^2-1) e^{K
\sum_n \sum_i s_n s_{n+i}} \Longrightarrow \prod_m \int_{-\infty}^{+\infty} ds_m e^{-\frac{1}{2} b
s_m^2 - u s_m^4} e^{K \sum_n \sum_i s_n s_{n+i}}\]
Where on the resulting expression on the right hand side we have smoothed the sharp delta function peaks into narrow Gaussian peaks (see figure~\ref{delta} ). 
Writing the above in fourior space with:
\[s_m = \int_q \sigma_\bq e^{i \bq.\bn},~~~~~~where~~~\int_q = \frac{1}{(2\pi)^d}\int_{-\pi}^{+\pi}
dq_1 ....\int_{-\pi}^{+\pi} dq_d,\]
transforms the partition function to:

\[Z=\prod_m \int_{-\infty}^{+\infty} ds_m e^{-\frac{1}{2}\int_q (K \sum_j
|e^{iq_j}-1|^2 +b-2Kd)\sigma_\bq\sigma_{-\bq} - u \int_{q_1}\int_{q_2}\int_{q_3} \sigma_{\bq_1}
\sigma_{\bq_2}\sigma_{\bq_3}\sigma_{-\bq_1-\bq_2-\bq_3}}.\]

The program is: {\it We will perform one quick step and then the sequence of steps! With this we will end up at two coupled algebraic equations. From there we can compute the critical indices.}

One quick step:

Let us assume that the atomic scale fluctuations do not effect the value of the critical indices
(an assumption which can only be verified a-posterior). Thus we will consider wavelengths much larger than the lattice constant. A momentum cut-off
$|\bq|<q_c$ will be imposed in the integrals (where $q_c<< \frac{2\pi}{a},~~a=1$, $a$ is the lattice
constant). Thus one can approximate $|e^{i q_i}-1|^2\simeq q_i^2$. But limiting $|\bq|<q_c$ harms: \[s_m=\int_q \sigma_\bq e^{i\bq.\bn}\]
One cannot get back the original $s_m$ (as large momenta $|q|>q_c$ are excluded in the limits of
the integral!)  So there is a technical problem. To resolve this one is forced to define the functional integral to count all possible states that the system can realize:
\[Z=\fis e^{-\frac{1}{2}\int_q (q^2+r)\sigma_\bq\sigma_{-\bq} - u \int_{q_1}\int_{q_2}\int_{q_3}
\sigma_{\bq_1}
\sigma_{\bq_2}\sigma_{\bq_3}\sigma_{-\bq_1-\bq_2-\bq_3}},~~~~~~~r=\frac{b- 2 K d}{K}. \]
Where
\[\fis = \prod_j \int_{-\infty}^{+\infty}(....)d\sigma_{q_j}\]
And now \[\int_q = \frac{1}{(2 q_c)^d}\int_{-q_c}^{+q_c}
dq_1 ....\int_{-q_c}^{+q_c} dq_d.\]
This is one typical example of defining an Effective Field Theory (EFT)! Our aim is to execute Kadanoff's idea of ``thinning out the degrees-of-freedom''. Here we are doing in momentum space (instead in real space as was done in the previous example of the last subsection). Higher momentum states will be integrated out {\it iteratively}. The logic of the procedure is given below. The fundamental importance of Wilson's iterative procedure (the renormalization group) is that we will get the pivotal function $f(K)$ of the previous example. From that it is possible to get the critical point and all the critical indices etc.

The Procedure: Define the functional integral as: $Z =\fis e^{H[\sigma]}$ with $H[\sigma] = -\frac{1}{2}\int_q
(q^2+r)\sigma_\bq\sigma_{-\bq} - u \int_{q_1}\int_{q_2}\int_{q_3}
\sigma_{\bq_1} \sigma_{\bq_2}\sigma_{\bq_3}\sigma_{-\bq_1-\bq_2-\bq_3} $. This functional integral can be written as the product of two functional integrals (as is clear from the definition of the functional integral above).
\[Z = \fiszero \underbrace{\left(\fisone e^{H[\sigma_0+\sigma_1]}\right)}_{higher~momenta~
integrals}\]
Here 
\[\sigma_\bq = \left\{
\begin{array}{l l}
\sigma_{0,\bq} \equiv \sigma_0 &  if ~ 0<|\bq|<\frac{q_c}{2}\\
\sigma_{1,\bq} \equiv \sigma_1 &  if~ \frac{q_c}{2}<|\bq|<q_c
\end{array} \right.\]
We want to integrate out $\sigma_1$s (the higher momentum shell). Also rename various parts of the Hamiltonian as:
\[H[\sigma] = \underbrace{-\frac{1}{2}\int_q
(q^2+r)\sigma_\bq\sigma_{-\bq}}_{H_F[\sigma]} + \underbrace{- u \int_{q_1}\int_{q_2}\int_{q_3}
\sigma_{\bq_1} \sigma_{\bq_2}\sigma_{\bq_3}\sigma_{-\bq_1-\bq_2-\bq_3}}_{H_I[\sigma]}.\]
Let the integration of higher momenta states yields:
\[e^{H^\p[\sigma^\p]} = e^{H_F[\sigma_0]} \fisone e^{H_F[\sigma_1] + H_I[\sigma_0+\sigma_1]}.\]

Here we define $\sigma^\p \equiv \sigma^\p_{\bq^\p}= \frac{1}{\zeta}\sigma_{0,\bq}$ (i.e., $\sigma^\p$ and $\bq^\p$ are rescaled
variables with $0<|q^\prime|<q_c$ and $0<|q|<\frac{q_c}{2}$).
$H_F[\sigma_0]$ can be written in terms of scaled variables trivially as:
\[-\frac{1}{2}\int_{0<|\bq|<\frac{q_c}{2}} (q^2 +r) \sigma_\bq\sigma_{-\bq} \Longrightarrow
-\frac{1}{2} (\zeta^2 2^{-d-2})\int_{0<|\bq^\p|<q_c} \underbrace{({q^\p}^2 + 4
r)}_{notice!} {\sigma^\p}_{\bq^\p} \sigma^\p_{-\bq^\p}.\] This trivial scaling cannot be done on the interaction term (due to coupling of momenta). One must use  perturbation theory and express:
\[e^{H_I[\sigma]} \simeq 1 - \times +\frac{1}{2} (\times)(\times)-... \]
Where $\times\equiv -H_I[\sigma]$. In this series expansion inside the functional integral one will encounter terms like:
\[I(\bq_1,\bq_2,....\bq_k) = \fisone \sigma_{1,\bq_1}\sigma_{1,\bq_2}.........\sigma_{1,\bq_k}
e^{H_F[\sigma_1]} \propto  \left\{\sigma_{1,\bq_1}\sigma_{1,\bq_2}
\sigma_{1,\bq_3}\sigma_{1,\bq_4} ......+ ....all~combinations... \right\}.\]
For the proof of it see the original article: Wilson-Kogut in\cite{2}. After doing the functional integral over $\sigma_1$s  (say up to 2nd order) one must rescale the $\sigma_0$ as before (by expanding the momenta $0<|\bq|<\frac{q_c}{2}$ to $0<|\bq^\p|<q_c$ etc). Effective Hamiltonian (after integrating upto second order (see for details Wilson-Kogut in\cite{2})) takes the form:
\[H[\sigma^\p] = \underbrace{-\frac{1}{2}\int_{q^\p}
({q^\p}^2+r^\p)\sigma^\p_{\bq^\p}\sigma^\p_{-\bq^\p}}_{H_F[\sigma^\p]} + \underbrace{- u^\p
\int_{q^\p_1}\int_{q^\p_2}\int_{q^\p_3}
\sigma^\p_{\bq^\p_1}
\sigma^\p_{\bq^\p_2}\sigma^\p_{\bq^\p_3}\sigma^\p_{-\bq^\p_1-\bq^\p_2-\bq^\p_3}}_{H_I[\sigma^\p]}.\]
With
\[r^\p =4 (r+3 c \frac{u}{1+r})\]
\[u^\p = 2^{4-d} (u - 9 c \frac{u^2}{(1+r)^2}).\]
\begin{figure}
\centering
\begin{tabular}{cc}
\includegraphics[height = 4cm, width = 6cm]{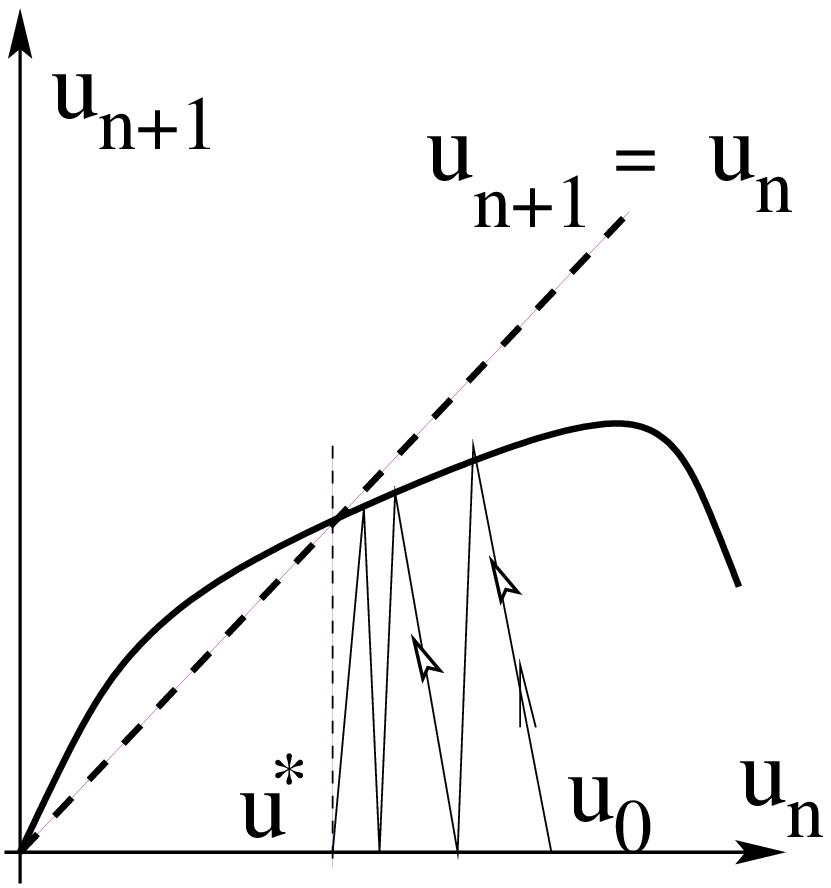}&
\includegraphics[height = 4cm, width = 6cm]{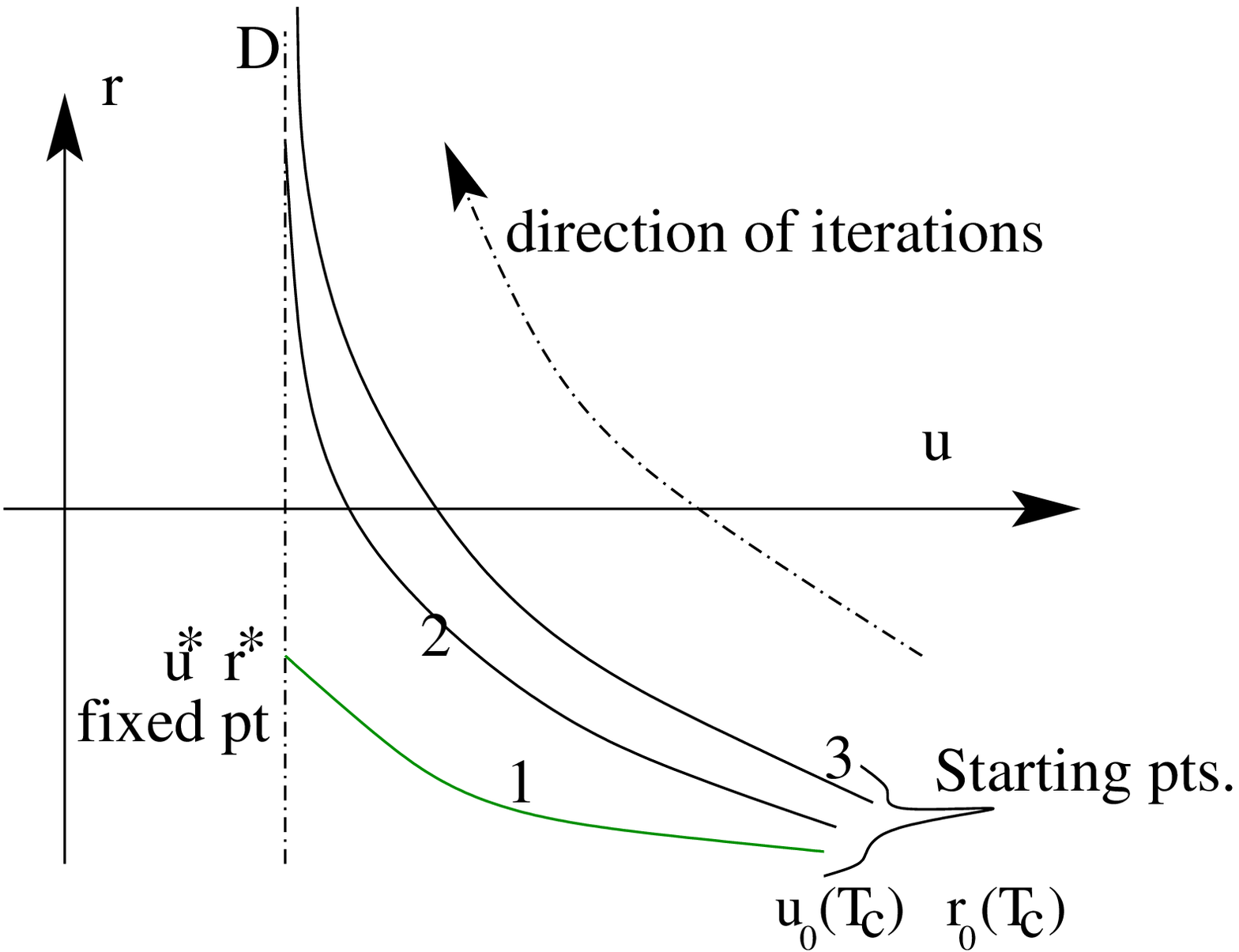}\\
(a)&(b)
\end{tabular}
\caption{(a)Iteration formula for $u_n$. (b) Calculation of $\nu$: if one starts with some initial $u_0,r_0$ at $T=T_c$, after sufficient number of iterations $r\rta r^\ast, ~~u\rta u^\ast$ (the fixed point). If one starts at $T \neq T_c$ iterations take the initial conditions to some asymptotic line $D$.}
\label{iter}
\end{figure}

These are the coupled algebraic equations. One can make analogy to the equation $K_m^{eff} = f(K_{m-1}^{eff})$ that we derived by using Kadanoff's real space blocking method. Here calculation is done in momentum space.  By invoking the same logic that at the critical point the function $f$ becomes scale independent due to the divergence of correlation length. This divergence of correlation length leads to fixed point solution (as we learn in Kadanoff's method in last subsection).  Thus self-consistent solution of the above contains the physics of the critical point.  One can obtain the self consistent solution by iterative procedure i.e., by evolving the above equations as a recursive procedure. This iterative procedure can be depicted by the following diagram (figure ~\ref{iter}(a)). Let us express the equations after nth iteration as
\[r_{n+1} =4 (r_n+3 c \frac{u_n}{1+r_n})\]
\[u_{n+1} = 2^{4-d} (u_n - 9 c \frac{u_n^2}{(1+r_n)^2}).\]
After sufficient number of iterations one will end up at the critical point $u^\ast,~r^\ast$. For illustration  consider the case of $u_n$ (figure \ref{iter}(a)). The two curves (one for $u$ and other for $f(u)$) meet at point $u^\ast$. If one starts with some initial value $u_0$ ($u_0>u^\ast$ or $u_0<u^\ast$, does not matter) one will end up, after sufficient number of iteration, at $u^\ast$. 

 {\it The great advantage of the Wilson's formulation is that we are able to know the required function explicitly.}  The other  important point is that the critical fixed point $r^\ast,~u^\ast$ does not depend on the initial parameters in the Hamiltonian i.e., $r_0,~u_0$. {\it Thus the critical point behavior does not depend on the microscopic details of the system. This is called the universality. For example, $\beta=1/3$ both for the magnetic systems and for the liquid gas transitions!(see for details \cite{2,3,4,10,26}).}

This how Wilson changed Kadanoff's qualitative ideas into a powerful computational procedure, as we know today, RG has been applied to wide variety of problems. Here by considering one example we have covered the basics  of the method.  

One can also compute the critical indices. Let us take one example of the calculation of $\nu$. For this we need to linearize the fixed point equations. For $T$ very near to $T_c$ (say curves 2, or 3 in figure ~\ref{iter}(b)):
\[r_l(T) = r_l(T_c) + (T-T_c)\frac{d r_l}{d T}|_{T_c}+...\]
\[u_l(T) = u_l(T_c) + (T-T_c)\frac{du_l}{dT}|_{T_c}+...\]
As $r= \frac{b- 2 K d}{K}$ is a function of temperature through $K=\frac{J}{k_B T}$, and $u=\frac{U}{k_B T}$ is also a function of temperature. Thus these can be Taylor expanded for $T$ near $T_c$. For very large $l$:
\[r_l(T) = r^\ast + (T-T_c)\frac{d r_l}{d T}|_{T_c}+...\]
\[u_l(T) = u^\ast + (T-T_c)\frac{du_l}{dT}|_{T_c}+...\]
Or
\[r_{l+1}(T) - r^\ast = (T-T_c)\frac{d r_{l+1}}{d T}|_{T_c}+...\]
\[u_{l+1}(T) - u^\ast = (T-T_c)\frac{du_{l+1}}{dT}|_{T_c}+...\]

For $T$ sufficiently close to $T_c$. As $r_{l+1} = f(r_l,u_l)$ and $u_{l+1} = g(r_l,u_l)$ (our fixed point equations), the temperature derivatives can again be Taylor expanded around the fixed point. This leads to:
\[r_{l+1} - r^\ast = c_1 (r_l-r^\ast) + c_2 (u_l-u^\ast)\]
\[u_{l+1} - u^\ast = c_3 (r_l-r^\ast) + c_4 (u_l-u^\ast)\]
As the functions $f(r_l,u_l)$ and $g(r_l,u_l)$ are known one can work out the expansion coefficients explicitly (see for details Wilson-Kogut in \cite{2}):
 \[\left(
\begin{array}{l}
   r_{l+1}-r^\ast \\
    u_{l+1}-u^\ast
  \end{array}
\right) = M_{2\times2}
\left(
\begin{array}{l}
   r_{l}-r^\ast \\
    u_{l}-u^\ast
  \end{array}
\right)
\]
By iterating $n-$times:
 \[\left(
\begin{array}{l}
   r_{l+n}-r^\ast \\
    u_{l+n}-u^\ast
  \end{array}
\right) = (M_{2\times2})^n
\left(
\begin{array}{l}
   r_{l}-r^\ast \\
    u_{l}-u^\ast
  \end{array}
\right)
\]
The purpose of iterating $n$-times is that the matrix $M^n$ is completely dominated by the largest eigenvalue  $\lambda^n$ ($\lambda=4$ for $u=0$ case). One can find eigensystem of the above matrix and finally (as the largest eigenvalue dominates!):
\[r_{l+n} - r^\ast \simeq \lambda_1^n w_{11} c_l (T-T_c).\]
here $\lambda_1$ is the largest eigen value, $w_{11}$ is the corresponding eigenvector, and $c_l$ is a constant (see for details Wilson-Kogut in\cite{2}). And a similar expression holds for the $u_l$. One observe that:
\[r_{l+n+1}(T) -r^\ast~~at~~(T-T_c =\frac{\tau}{\lambda_1}) = r_{l+n}(T)-r^\ast~~at~~T-T_c =\tau\]
Thus the correlation length:
\[\xi(r_{l+n+1},u_{l+n+1})\left|_{T=T_c+\tau/\lambda_1} = \xi(r_{l+n},u_{l+n})\right|_{T=T_c
+\tau}.\]
Now
\[\frac{1}{2^{l+n+1}}\xi(T_c+\tau/\lambda_1) = \frac{1}{2^{l+n}}
\xi(T_c+\tau),~~~~~as~~~\xi(r_{l+n},u_{l+n})=\frac{1}{2^{l+n}} \xi(r_0,u_0).\]
If correlation length scales as: \[\xi(T_c+\tau) \propto \tau^{-\nu}\]
Then from above:
\[(\frac{\tau}{\lambda_1})^{-\nu} = 2 \tau^{-\nu} \Longrightarrow \nu=\frac{ln2}{ln\lambda_1}.\]
 If $u=0$ (simple Gaussian model), then $\lambda_1=4$ which leads to $\nu =0.5$ (the MFT result). With explicit calculation of $\lambda_1$ for $u\ne 0$ (see for details Wilson-Kogut in\cite{2}), $\nu$ comes out to be:
\[\nu \simeq 0.58\]
This is very close to the experimentally observed value $\nu = 0.6$ (in MFT it is
$0.5$!!!!). Similarly other critical indices can be calculated and are quite close to what is observed experimentally\cite{10,26}.

\section{Why do MFTs fail and how does RGT rectify that?}

Thus, by exploiting scale invariance at the critical point and by finding the critical point from fixed point equations by iterative procedure one is able to calculated correct values of critical indices.  The important point that goes implicit in the above statement is that iterative procedure is not only a mathematical device to reach to the fixed point but also at each step of iteration one is averaging over smaller length scale (or higher momentum scale) fluctuations and with this appropriate averaging over increasingly longer length scales one is able calculate the correct values of critical indices. With this sophistication of renormalization group method we immediately see the drawback of MFTs. In MFTs we do an abrupt step in dealing with the fluctuations: 
$\la\sigma\ra = \tanh(h + \frac{T_c}{T}\la\sigma\ra)$
i.e., we average out only the smaller length scale (of the order of lattice constant) fluctuations (in calculating the effective field seen by the test spin). Near the critical point just below the critical temperature, ensemble averages over smaller length scales (of the order of lattice constant) lead to zero value of the order parameter (here mean magnetization). Only when one averages over a large length scale (of the order of correlation length) can one see some tiny magnetization just below criticality. Thus our assumption of tiny mean magnetization over a smaller length scale (of the order of lattice constant) in MFT breaks down! And MFT predicts wrong values of critical indices( see also\cite{6}).

\section{Appendix: K. G. Wilson--a short biographical sketch}
\begin{figure}[h!]
\includegraphics[height = 4cm, width = 6cm]{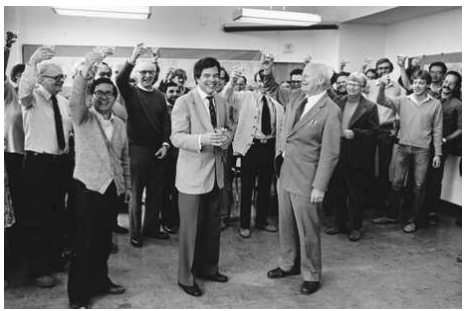}
\caption{Kenneth Geddes Wilson (8 June 1936 -- 15 June 2013) with Hans Bethe and colleagues; 1982 Nobel
Prize in Physics for the renormalization group method (Image source: http://news.cornell.edu/stories/2013/06/physics-nobel-laureate-kenneth-wilson-dies).}
\end{figure}
Ken Wilson earned his Ph.D. from the California Institute of Technology (CalTech) in 1961, studying under Murray Gell-Maan. Elementary Particle Physics (now called high energy physics) was a fashionable subject in those days (today also it is one of the mainstream  topics along with condensed matter physics) and students would like to opt for it. But Wilson rebelled, and worked for a while on
plasma physics. Something that did not go very well\cite{2}. During that time he used to discuss his problems with S. Chandrasekhar (cousin of sir C. V. Raman). Here is an interesting story\cite{11}:

His father had told him: 'When you go to Caltech, be sure to meet the two great physicists there
namely, Feynman and Gell-Mann'. K. Wilson says that when he called on Feynman, the latter seemed to
be gazing at the ceiling. He asked Feynman, 'Professor, what are you working on at
present?' Feynman replied, 'Nothing! Wilson then went to Gell-Mann and became his student!!!

He worked on the solutions of Gell-Mann and Low equation $\lambda ^2 \frac{d(e_\lambda^2)}{d
\lambda^2} = \phi(e_\lambda^2,m^2/\lambda^2)$ and on fixed-source meson theory for his PhD Thesis.
Around 1963 he read Onsager's paper and became interested in critical phenomena, and after that studying other works on critical phenomena he 
 developed the renormalization group theory, using tricks and techniques that he had learnt in particle physics.

\end{document}